\documentclass{aastex62}
\usepackage{hyperref}
\turnoffedit
\received{May 1, 2018}
\revised{Sep 18, 2018}
\accepted{Dec 15, 2018}
\submitjournal{ApJ}

\shorttitle{Multiple Spiral Arms in HD~34700A}
\shortauthors{Monnier et al.}
\begin{document}


\title{Multiple spiral arms in the disk around intermediate-mass binary HD 34700A}

\correspondingauthor{John Monnier}
\email{monnier@umich.edu}

\author[0000-0002-3380-3307]{John D. Monnier}
\affil{Astronomy Department, University of Michigan,
Ann Arbor, MI 48109, USA}

\author{Tim J. Harries}
\affil{Department of Physics and Astronomy, University of Exeter, Exeter, EX4 4QL, UK}

\author{Jaehan Bae}
\affil{Department of Terrestrial Magnetism, Carnegie Institution for Science, 5241 Broad Branch Road, NW, Washington, DC 20015, USA}

\author[0000-0001-5980-0246]{Benjamin R. Setterholm}
\affil{Astronomy Department, University of Michigan,
Ann Arbor, MI 48109, USA}

\author{Anna Laws}
\affil{University of Exeter, School of Physics of Astronomy, Astrophysics Group, Exeter, EX4 4QL, UK}

\author{Alicia Aarnio}
\affil{University of Colorado Boulder, Boulder CO USA 80303}

\author{Fred C. Adams}
\affil{Astronomy Department, University of Michigan,
Ann Arbor, MI 48109, USA}
\affil{Physics Department, University of Michigan,
Ann Arbor, MI 48109, USA}

\author{Sean Andrews}
\affil{Harvard-Smithsonian Center for Astrophysics, Cambridge, MA 91023 USA}

\author{Nuria Calvet}
\affil{Astronomy Department, University of Michigan,
Ann Arbor, MI 48109, USA}
\author{Catherine Espaillat}
\affil{Boston University, Boston, MA}

\author{Lee Hartmann}
\affil{Astronomy Department, University of Michigan,
Ann Arbor, MI 48109, USA}

\author{Stefan Kraus}
\affil{University of Exeter, School of Physics of Astronomy, Astrophysics Group, Exeter, EX4 4QL, UK}

\author{Melissa McClure}
\affil{European Southern Observatory, Garching, Germany}

\author{Chris Miller}
\affil{Astronomy Department, University of Michigan,
Ann Arbor, MI 48109, USA}

\author{Rebecca Oppenheimer}
\affil{American Museum of Natural History, New York, NY USA}

\author{David Wilner}
\affil{Harvard-Smithsonian Center for Astrophysics, Cambridge, MA 91023 USA}

\author{Zhaohuan Zhu}
\affil{University of Nevada, Las Vegas}

\begin{abstract}

We present the first images of the transition disk around the close binary system HD~34700A in polarized scattered light using the Gemini Planet Imager instrument on Gemini South. The J and H band images reveal multiple spiral-arm structures outside a large ($R=0.49\arcsec = 175$\,au) cavity along with a bluish spiral structure inside the cavity.  The cavity wall shows a strong discontinuity and we clearly see significant non-azimuthal polarization $U_\phi$ consistent with multiple scattering within a disk at an inferred inclination $\sim$42$\arcdeg$. Radiative transfer modeling along with a new {\em Gaia} distance suggest HD~37400A is a young ($\sim$5~Myr) system consisting of two intermediate-mass ($\sim$2\,M$_\odot$) stars surrounded by a transitional disk
and not a solar-mass binary with a debris disk as previously classified.  Conventional assumptions of the dust-to-gas ratio would rule out a gravitational instability origin to the spirals while hydrodynamical models using the known external companion or a hypothetical massive protoplanet in the cavity both have trouble reproducing the relatively large spiral arm pitch angles ($\sim30\arcdeg$) without fine tuning of gas temperature.  We explore the possibility that material surrounding a massive protoplanet could explain the rim discontinuity after also considering effects of shadowing by an inner disk.  Analysis of archival Hubble Space Telescope data suggests the disk is rotating counterclockwise as expected from the spiral arm structure and  {revealed a new low-mass companion at 6.45\arcsec separation.  We include an appendix which sets out clear definitions of Q, U, Q$_\phi$, U$_\phi$, correcting some confusion and errors in the literature.
 }

\end{abstract}



\vspace{.2in}
\section{Introduction} 
\label{sec:intro}
Planet formation relies on the interplay of several physical processes involving dust, ice, gas, chemistry, as well as the radiation field from the central star as shadowed by inner disk structures.  Observations are needed to determine the importance of effects such as gravitational instability \citep{boss1997}, streaming instability \citep{johansen2007}, dust growth \citep{birnstiel2010}, core accretion \citep{pollack1996}, planetary migration \citep{tanaka2002}, and more.  Theorists hope to eventually build a predictive framework that can explain the observed demographics of exoplanets around low- and intermediate-mass stars, but we are currently far from achieving this goal.

Fortunately, modern high angular resolution techniques have opened powerful new ways to validate physical models. For low mass stars, mm-wave imaging \citep[e.g.,][]{fedele2018,huang2018} and scattered-light coronographic imaging \citep[e.g.,][]{rapson2015a,avenhaus2018} routinely find symmetric ring structures possibly caused by accreting or still-forming protoplanets \citep{bae2017}.  For intermediate-mass (1.5-3\,M$\odot$) stars, we find more varied structures, such as asymmetric complex disks \citep[e.g., AB Aur;][]{oppenheimer2008}, spirals \citep[e.g., MWC758, HD135344B, HD142527;][]{garufi2017} in addition to multiple rings \citep[e.g., HD163296, HD169142;][]{monnier2017}.  
\citet{avenhaus2018} pointed out that spiral structures appear mainly around intermediate-mass stars and not the lower-mass T Tauri stars.  The explanation for this dichotomy is not known, but larger stars tend to have higher disk masses and higher companion fractions, both of which lead to more spiral  structure.    

In this work, we present the discovery of one of the most ``spiral-armed'' disks so far around HD~34700A, with structures reminiscent of the HD~142527 system \citep{avenhaus2017}.  HD~34700A is a close binary system (period 23.5 days) originally thought to consist of two nearly equal-mass main-sequence solar-mass stars (spectral type G0, $T_{\rm eff}\sim6000$K) at 125pc with large far-infrared excess interpreted as a ``Vega-like'' (debris) disk \citep{torres2004}.  {\citet{torres2004} anticipated that a farther distance would mean a more massive and younger system and indeed the new {\em Gaia} distance now places this system at 356.5$\pm$6.1\,pc, nearly 3 times farther away than previously assumed.} {Assuming solar metallicity and the new distance, we find HD~34700A to consist of two $\sim$2.05\,M$_\odot$ stars with nearly identical effective temperatures (5900K \& 5800K) and a system age of $\sim5$\,Myrs (details provided in \S\ref{stars}). We can now interpret the infrared excess as a transition disk with ongoing planet-formation rather than an older, more evolved ``debris'' disk .}

The closest known stellar companion (HD~34700B) to the inner pair of stars (HD~34700Aa,Ab) was reported by \citet{sterzik2005} at separation of 5.18$\arcsec$ (projected separation of 1850 au) at PA 69.1$^\circ$ with photometry $J=12.29$, $H=11.52$, and $K=11.03$.  Assuming a coeval system and that the K band flux is entirely unreddened stellar flux, this suggests that the HD~34700B is a 0.7 $M_\odot$ K7 star with $T_{\rm eff}=4000$K \citep{siess2000}. { {\em Gaia} DR2 now includes this companion and it seems to be at the same distance at HD 34700A.}  We will discuss the possibility of tidal interactions of HD~34700B with the HD~34700A disk later in this paper.  \citet{sterzik2005} report a slightly-fainter fourth component at $9.2\arcsec$ (projected separation of 3300\,au; {listed but with no parallax solution yet in {\em Gaia} DR2}) which we will not discuss further although it could potentially also play a role in sculpting the outer disk structure of HD~34700A depending on the true 3-D orbital geometry.

After describing our new GPI observations in detail, we present a simple radiative transfer model that explains many of the observed properties of the disk, although serious deficiencies remain.  Lastly, we ran hydrodynamical simulations tuned for HD 34700A and discuss the difficulty in matching the large pitch angle spirals with conventional disk prescriptions, for both an outer perturber (HD~34700B) or a protoplanet in the mostly dust-free cavity.  {In an appendix, we clearly define the Stokes conventions adopted here, correcting some confusion found in the literature. Also in an appendix, }
we present a preliminary analysis of archival HST data, identifying a possible fifth member of this system. ALMA data will be needed in order to allow a comprehensive modeling of the HD~34700A disk and to determine the physical origin of the extensive spiral structures.

\section{Observations and Data Processing}
\label{data}
We report new imaging of HD~34700A using the Gemini Planet Imager \citep[GPI;][]{macintosh2008,macintosh2014,poyneer2016} installed on Gemini South. In polarimetry mode \citep{perrin2015} with the adaptive optics system and an occulting spot, GPI can obtain high dynamic range imaging of scattered light from Y-K bands relying on the physics of scattering to deliver a  distinctive polarization pattern. Light scattered from dust grains {surrounding a star} will be polarized with E-field vectors aligned perpendicular to the radial direction toward the star, while the light from the central star's point spread function (PSF) will be typically unpolarized or linearly polarized throughout the PSF. 

We observed HD~34700A on UT 2018 January 3 utilizing the standard GPI coronagraphic configurations (specifically `J-coron' and `H-coron'), including use of a coronographic spot (0.184'' diameter for J band and 0.246'' diameter for H band) and appropriate Lyot and apodizing pupil masks. We chose $\sim$30s integration time to avoid detector saturation by light around the spot. We coadded 2 frames together to accumulate $\sim$1~minute of on-source exposure time per file, a limit imposed by the rotating field-of-view in the GPI design.  We used the Wollaston prism mode and rotated the half-wave plate 22.5\arcdeg~between {four 1~minute observations} to determine the Stokes parameters {(half-wave plate angles 0$\arcdeg$, 22.5$\arcdeg$, 45$\arcdeg$, 67.5$\arcdeg$ were used).  } {A total of 32 frames were saved for J and H band observations, leading to 8 independent sequences with 4 equally-spaced half-wave plate angles.} Table~\ref{table:targets} contains the information on the target star while Table~\ref{table:obslog} contains the Observing Log.  

{For this ``discovery'' paper, we have used only GPI pipeline primitives to simplify the data reduction description. 
All steps to create a flux-calibrated Stokes datacube (I,Q,U,V), {including bias correction, bad-pixel corrections, flat-fielding, and flux calibration,} were carried out using the IDL-based GPI pipeline version 1.4.0 (downloaded on 2018Aug06). The general calibration procedures have been described by \citet{perrin2014} and \citet{millar-blanchaer2016} with discussion of flux calibration by \cite{hung2016}. } 

\subsection{Analysis Steps}
{\noindent Here we give detail on each of the major analysis steps leading to the calibrated Stokes datacubes.}\\

{ {\bf Locating Star Position:} }
{
In order to precisely determine the star's location behind the coronagraphic spot and to calibrate the flux scale for GPI, the instrument contains a diffractive element in the pupil plane that creates so-called ``satellite spots.'' Each of these spots has radial structure that points back toward the location of the star and which contains a certain percentage of the flux from the star. The GPI pipeline centers each frame using a high-pass filter followed by a Radon transform to find the stellar location to within $\pm$0.1~pixels (pixel scale 14.14~milliarcseconds) using the symmetry of the diffractive spots \citep{wang2014}. Also, by carrying out aperture photometry on these spots, one can deduce the flux from the star assuming the photometry in Table~\ref{table:targets}.   The H-band spots in polarization mode were extensively studied by \citet{hung2016} and a $\pm$13\% systematic error was recommended to be used in addition to any statistical error due to variations in spot-to-star flux ratios observed during engineering studies.  The J-band polarization mode has not been studied as systematically -- here, we use the second-order satellite spots, which contain $\sim$25\% less flux\footnote{This and related GPI calibration information will be posted publicly on the Gemini Observatory website at http://www.gemini.edu/sciops/instruments/gpi/} than the first-order spots, because they are better separated from the speckle halo and adopt a 20\% systematic error (private communication,  Robert De Rosa, GPI team).  
}

{{\bf Calibrating Flux:}}
{
 We median-combined all the total intensity images in the instrument frame before estimating photometry and our errors combine systematic and statistical errors in quadrature.  We report calibration factors in the form following \citet{wolff2016}: 1 ADU/sec/coadd = $X$ mJy/arcsec$^2$, where X is the calibration factor.  For J-band we find calibration factor to be 3.7 mJy/(ADU/s)/coadd/arcsec$^2$ $\pm$ 22\% and H band we find 3.7 mJy/(ADU/s)/coadd/arcsec$^2$ $\pm$ 15\% (coincidentally, they are the same number within 2 significant figures),  We note that these values are only appropriate for photometry of point source detections and can not be strictly applied for diffuse, extended brightness distributions.  Proper photometry of the diffuse component requires knowledge of the PSF, including Strehl and scale of the residual speckle halo. }

{ {\bf Removing polarized flux from star PSF:}}
Because of the high level of scattered light from circumstellar dust -- seen even in individual frames -- we used only light behind the coronagraphic spot for estimating the  linear polarization of the stellar signal \citep[i.e.,][]{millar-blanchaer2016}.  Once we have the fractional $f_{Q,U,V}$ of light behind the spot, we can multiply this by the total intensity $I$ in each pixel throughout the PSF to estimate the $Q,U,V$ contamination and subtract these contributions from the linear polarization.  For reference, we report the mean stellar linear polarization ($P_{\rm band} = (f^2_Q + f^2_U)^\frac{1}{2}$, $\Theta = \frac{1}{2}\arctan{\frac{U}{Q}}$) that we removed (all angles are degrees East of North): 
{HD~34700A $P_J=0.61\%\pm0.08\%$ at $\Theta=12\arcdeg\pm5\arcdeg$,      
 $P_H=0.82\%\pm0.08\%$ at $\Theta=-17\arcdeg\pm8\arcdeg$.}
The errors reported above are based on the scatter amongst the 8 independent Stokes datacubes (each based on 4 HWP positions) and do not include systematics. The observed stellar polarization angle $\Theta$ varied $\sim10\arcdeg$ as a function of parallactic angle within each set, strongly suggesting that these measurements are partially contaminated by uncorrected instrumental effects and not purely intrinsic. We refer the reader to the GPI instrumentation papers referenced in \S\ref{data} for more information on the systematic errors related to removal of the instrumental signature in the pipeline.  That said, generally our values are broadly consistent with the small level of polarization measured in V band by previous workers. Specifically,
\citet{bhatt2000} reported no detectable linear polarization in V band, 
while \citet{oudmaijer2001} found V band polarization  P$_V = 0.35\pm0.06\%$ with $\Theta\sim28\pm5\arcdeg$, consistent with interstellar origin.

{ {\bf Creating final Stokes datacubes:}}
{
Following subtraction of the mean stellar polarization signal from the Stokes data cube (one for each group of 4 files), we then median-combined multiple Stokes datacubes spanning a range of parallactic angles} { after rotating images to be aligned with ``True North.'' The current pipeline determines the sky orientation based on the calibration of \citet{konopacky2014} and the systematic error is estimated to be $\pm$0.13\arcdeg \citep{derosa2015}. } {Lastly, we project the traditional Stokes $Q,U$ components (oriented relative to North/East) onto a local Stokes $Q_\phi,U_\phi$ based on the stellar position determined earlier in the processing.  In this procedure \citep[see derivation in][]{schmid2006,avenhaus2014,garufi2014,millar-blanchaer2016},  linear polarization vectors that are perpendicular to the line connecting the star to the pixel are positive in $Q_\phi$ space while radial polarization patterns are negative.  Similarly polarization vectors oriented $\pm$45$\arcdeg$ from this are found in the $U_\phi$ component. See appendix~\ref{appendix:stokes} for more detail on the definition of $Q,U,Q_\phi, U_\phi$ since misleading descriptions are common in the literature.    This local projection is very practical since single-scattering should be oriented around the stellar position and produce purely positive $Q_\phi$ signal, while noise can be both positive and negative. Furthermore,  miscalibrations (especially in the inner PSF halo) will produce residual systematic $U_\phi$ signal that can be used to assess data quality and guard against false conclusions.  That said, we find here strong $U_\phi$ signal that we identify as an astrophysical signature of an inclined disk, consistent with predictions of \citet{canovas2015} and  \citet{dong2016} who explored polarization signatures for optically-thick, more edge-on disks.
}

{  {\bf Bootstrapping Errors:}
All error analysis in work has been based on bootstrap sampling, where we randomly sample the 8 independent Stokes datacubes (with replacement) and calculate the median Stokes datacube for 100 bootstraps. From this median Stokes cube, we then calculate Q$_\phi$,U$_\phi$.   These 100 bootstrapped synthetic datasets are available throughout the later analysis steps allowing all derived quantities (such as radial profiles, fraction polarizations, etc) to have their errors properly determined. While there will be systematic errors un-monitored by this, at least effects from readnoise, photon noise, bias correction, and varying AO performance will be represented in the uncertainties presented here. }

\begin{table}
\centering
\caption{Target Information}
\label{table:targets}
\begin{tabular}{lll}
\hline
Names & HD~34700A, HIP 24855 & \\
RA (J2000) &~~$ 05^{\rm{h}}$ $19^{\rm{m}}$ $41\fs42$ & \citet{gaia2018}\\
Dec (J2000) & $+05\degr$ $38\arcmin$ ~$42\farcs 80$ & \citet{gaia2018} \\
R mag & ~8.80$\pm$0.06 & \citet{fujii2002} \\
J mag & 8.041$\pm$0.023 & \citet{2mass} \\
H mag & 7.706$\pm$0.023 & \citet{2mass} \\
K$_s$ mag & 7.482$\pm$0.024 & \citet{2mass} \\
Spectral Type & G0$+$G0 & \citet{mora2001},\citet{torres2004} \\
T$_{\rm eff}$ & 5900K$+$5800K & \citet{torres2004} \\
Binary\tablenotemark{a} Period (days) & $23.4877\pm0.0013$ & \citet{torres2004} \\
Distance (pc) & $356.5\pm6.1$ & \citet{gaia2018} \\
\hline
\end{tabular}
\tablenotetext{a}{In addition to unresolved spectroscopic companion with period 23.5 days, \citet{sterzik2005} noted fainter companions at 5.2\arcsec and 9.2\arcsec away.}
\end{table}

\begin{table}
\centering
\caption{Observing Log of Polarimetry Observations using Gemini Planet Imager. For these observations we used the default occulting spot, apodizer, and Lyot stop appropriate for the observing waveband. }
\label{table:obslog}
\begin{tabular}{llcccccc}
\hline
\colhead{UT Date} & \colhead{Target Name}   & \colhead{Filter} & \colhead{T$_{\rm int}$ (sec)}         & \colhead{N$_{\rm coadds}$} &  \colhead{N$_{\rm Frames}$\tablenotemark{a}} & \colhead{Airmass} & \colhead{RMS Wavefront Error (nm)}  \\
\hline
2018 January 03 & HD~34700A & J & 29.10 & 2 & 32 & 1.25$-$1.32 & 206$-$252 \\
2018 January 03 & HD~34700A & H & 29.10 & 2 & 32 & 1.23$-$1.24 & 201$-$252 \\
\hline
\end{tabular}
\tablenotetext{a}{Here we refer to the number of frames used in the data reduction, where a frame consists of N$_{\rm coadds}$ images coadded with individual exposures times of T$_{\rm int}$ seconds at a single half-wave plate position.   }
\end{table}

\begin{figure}
\centering
\includegraphics[width=6in]{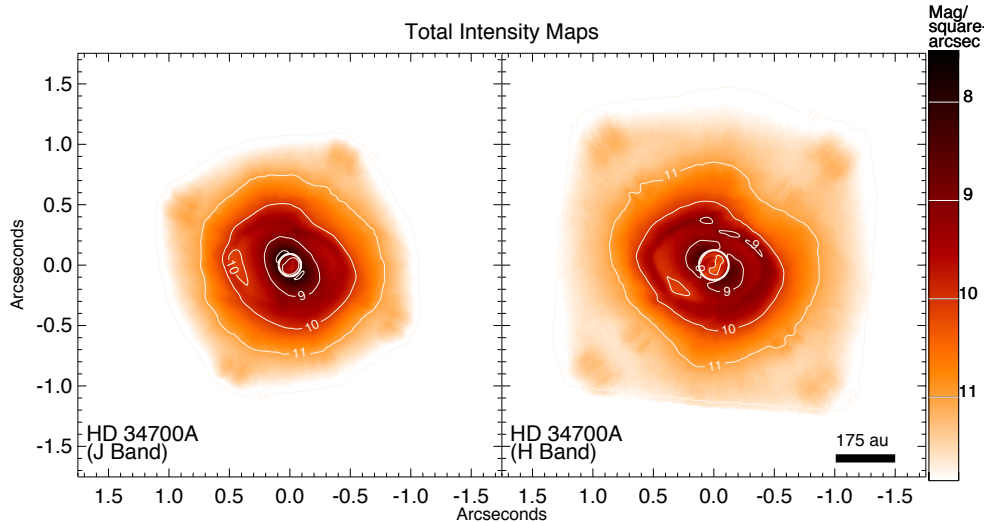}
\caption{Total intensity maps, measured by the Gemini Planet Imager, shown using a linear color table (see color bar).   The intensity scales with the local surface brightness levels, labeled with contours in units of Vega magnitudes / square-arcsecond.  The approximate location and size of the occulting spot is marked with a white circle in each panel.   East is left, North is up.  Note that most of the halo of light (especially within 0.3$\arcsec$) is from the residual stellar point spread function, although one can clearly see the ring-like circumstellar scattering from the HD~34700A transition disk.}
\label{fig:toti}
\end{figure}

\begin{figure}
\centering
\includegraphics[width=6in]{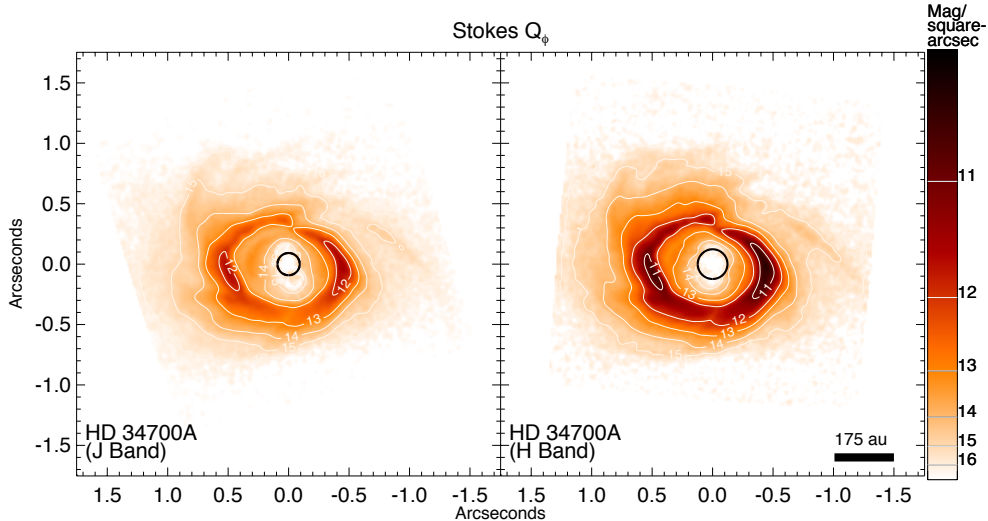}
\caption{$Q_\phi$ maps, measured using the Gemini Planet Imager, shown using a color table proportional to the square-root of absolute value of azimuthal-component of polarized intensity $Q_\phi$ (see text for description of this quantity).  The maps were smoothed by a {flux-conserving} Gaussian with FWHM 30~milliarcseconds (2.1~pixels) to improve SNR. The local surface brightness levels can be found as labeled contours in units of Vega magnitudes / square-arcsecond.  The approximate location and size of the occulting spot is marked with a black circle in each panel. East is left, North is up.  }
\label{fig:qr}
\end{figure}

\begin{figure}
\centering
\includegraphics[width=6in]{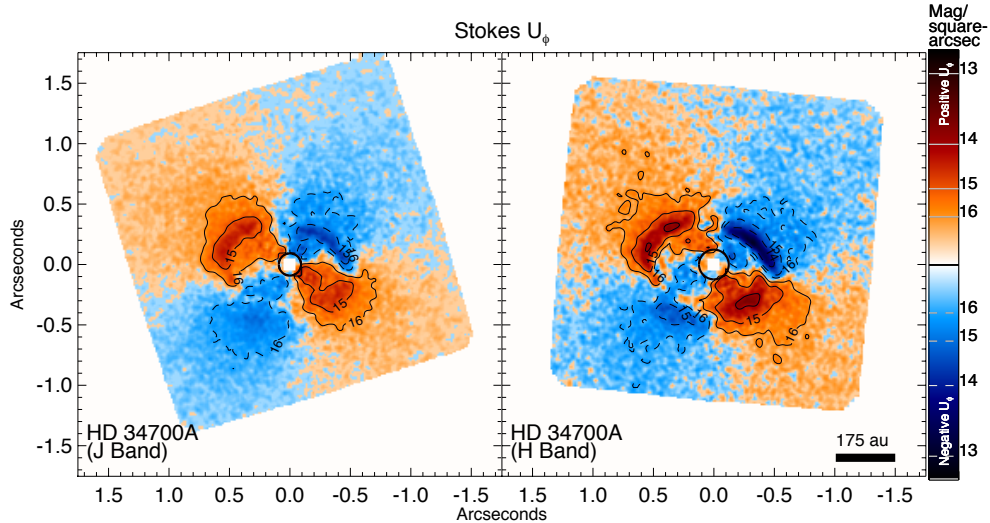}
\caption{$U_\phi$ maps, measured using the Gemini Planet Imager, shown using a color table proportional to the square-root of absolute value of polarized intensity $U_\phi$ (see text for description of this quantity), where red color shows positive $U_\phi$ and blue color shows negative $U_\phi$.  The maps were smoothed by a Gaussian with FWHM 30~milliarcseconds (2.1~pixels). The local surface brightness levels can be found as labeled contours in units of Vega magnitudes / square-arcsecond.  The approximate location and size of the occulting spot is marked with a black circle in each panel. East is left, North is up.  }
\label{fig:ur}
\end{figure}

\begin{figure}
\centering
\includegraphics[width=6in]{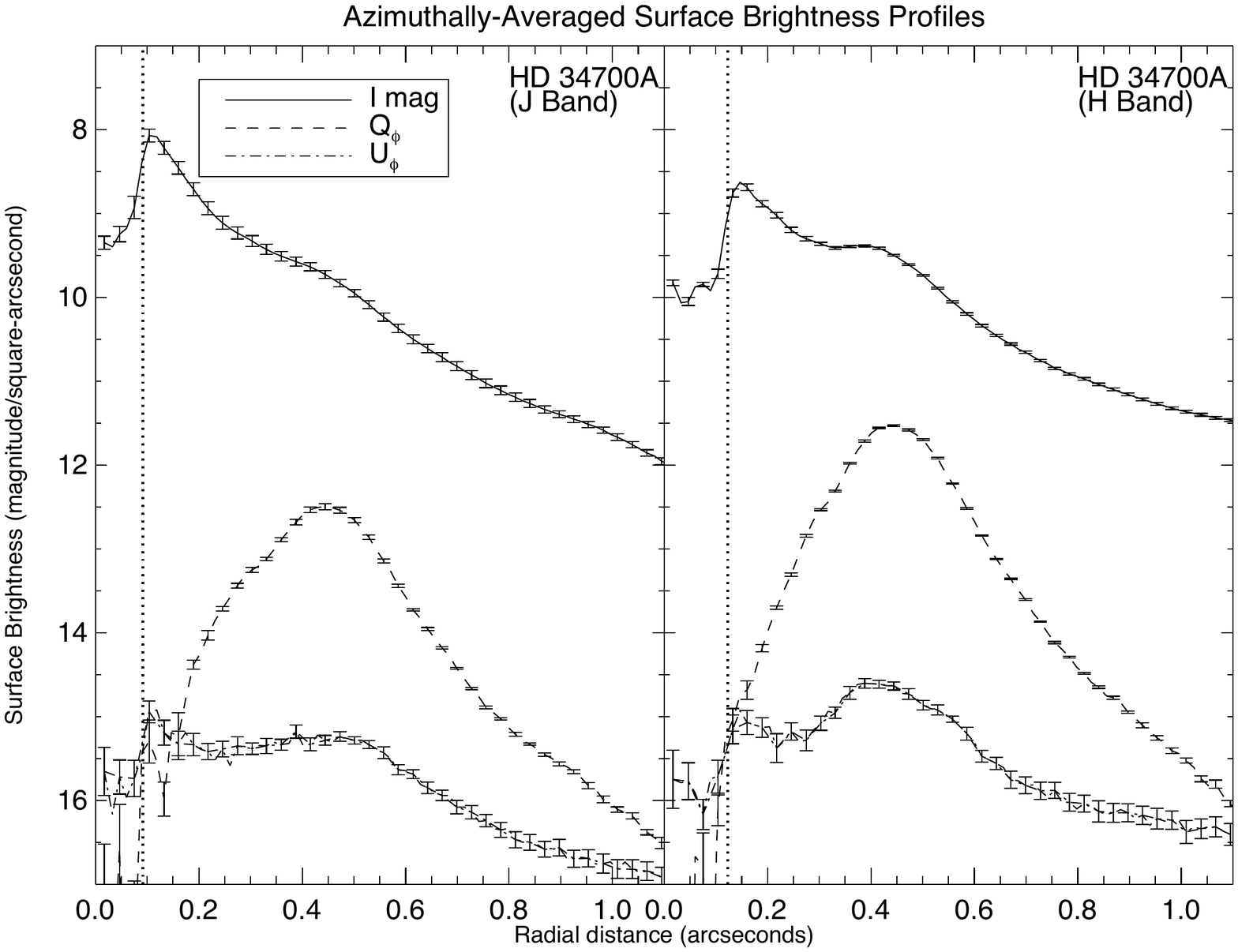}
\caption{Azimuthally-averaged surface brightness profiles for the J band and H band imaging of HD~34700A are shown here.
For total intensity and $Q_\phi$, the surface brightnesses were averaged in annuli centered on the stellar position, then the absolute value before plotting the logarithm. For $U_\phi$, the absolute value was taken first before averaging to avoid self-cancellation.  The dotted vertical line shows the radius of the coronagraphic spot used.  Errors were determined from bootstrap sampling of the 8 independent Stokes datacubes.  There is an additional overall flux calibration uncertainty (not shown) estimated to be $\pm$22\% at J band and $\pm$15\% at H band.
}
\label{fig:profiles}
\end{figure}

{
\subsection{Analysis Discussion}
}
{
There are some additional calibration steps that are often carried out in analysis of polarization data, most of which were expertly discussed in \cite{avenhaus2018}. Here is a brief list of calibration steps we did not include and our reasons.}

\begin{itemize}
\item {Methods that minimize $U_\phi$ using extra free parameters \citep[see][]{avenhaus2018}.  When looking at very faint polarized light, it is sensible to adjust free parameters representing instrumental calibration factors to minimize the amplitude of $U_\phi$ since this quantity is often intrinsically small or zero for single-scattering disks.  However, in our case we have very strong scattered light for a inclined disk and so it is not safe to assume $U_\phi$ is zero.  { \em It is by design that we do not attempt to minimize $U_\phi$ since such procedures may remove actual astrophysical signal.} }
\item {Deconvolution Methods. \citet{avenhaus2018} has an excellent demonstration of how the telescope point spread function will convolve the observed Q,U images, which can corrupt the $Q_\phi$ and $U_\phi$, especially too close to the star's position or if there are sharp changes in the intensity (i.e., near strong asymmetric features).  {In some cases, the local polarized signal level can be reduced compared to the true value.} {VLT-SPHERE observers have the option of collecting a quick PSF for each observation using an ND filter, but this option is not available for GPI and so we do not have an accurate PSF to allow for a deconvolution analysis.  The best we can do is to forward convolve our modeling to see what distortions in $Q_\phi$ and $U_\phi$ might be occur -- note we find none of this important for this star since the ring is well-resolved and not located near the coronographic spot.}}
\end{itemize}
~                

\section{Imaging Results for HD~34700A}
\subsection{Basic Description}
We present the total intensity maps in Figure~\ref{fig:toti}, Stokes $Q_\phi$ maps in Figure~\ref{fig:qr}, Stokes $U_\phi$ maps in Figure~\ref{fig:ur}, and their corresponding mean radial profiles in Figure~\ref{fig:profiles}.  Each figure has an explanation of how the images are scaled and presented.  We generally present color tables that are proportional to the square-root of intensity for higher dynamic range in order to see faint details in the outer disk.

First we discuss the total intensity maps in Figure~\ref{fig:toti}.  We see a depression in the center of the PSF because of the occulting mask, marked by a circle.  We see the inner PSF was elongated likely due to telescope wind-shake.  There are fuzzy spots outside the main PSF due to either residual ``waffle mode'' from the adaptive optics system or the diffractive satellite spots induced by GPI for registration of the bright star behind the coronagraph. We can clearly see the scattered light from the main dust ring here without using differential polarization -- we will be able to extract a crude scattering total intensity from the dust in this ring later in this paper.

Figure~\ref{fig:qr} contains the Stokes $Q_\phi$ maps, showing more detail of the large dust ring, including spiral arms structures.  It is useful to compare these images to the $U_\phi$ images in Figure~\ref{fig:ur} since residuals in the $U_\phi$ map often indicate the level of systematic errors in our analysis, either due to problems in the pipeline calibration or possibly effects of multiple scattering for edge-on systems \citep[as discussed by][]{canovas2015}.   Many workers analyzing polarization data adjust the pipeline calibration to minimize the butterfly pattern seen in $U_\phi$ under the assumption there is no astrophysical signature present -- this assumption should be tested for simulated disks  to see if such adjustments have the possibility to erase true signal.  In our case, radiative transfer modeling will support the conclusion that the bulk of our $U_\phi$ signal is real and not an instrumental artifact.

\begin{figure}
\centering
\includegraphics[width=6in]{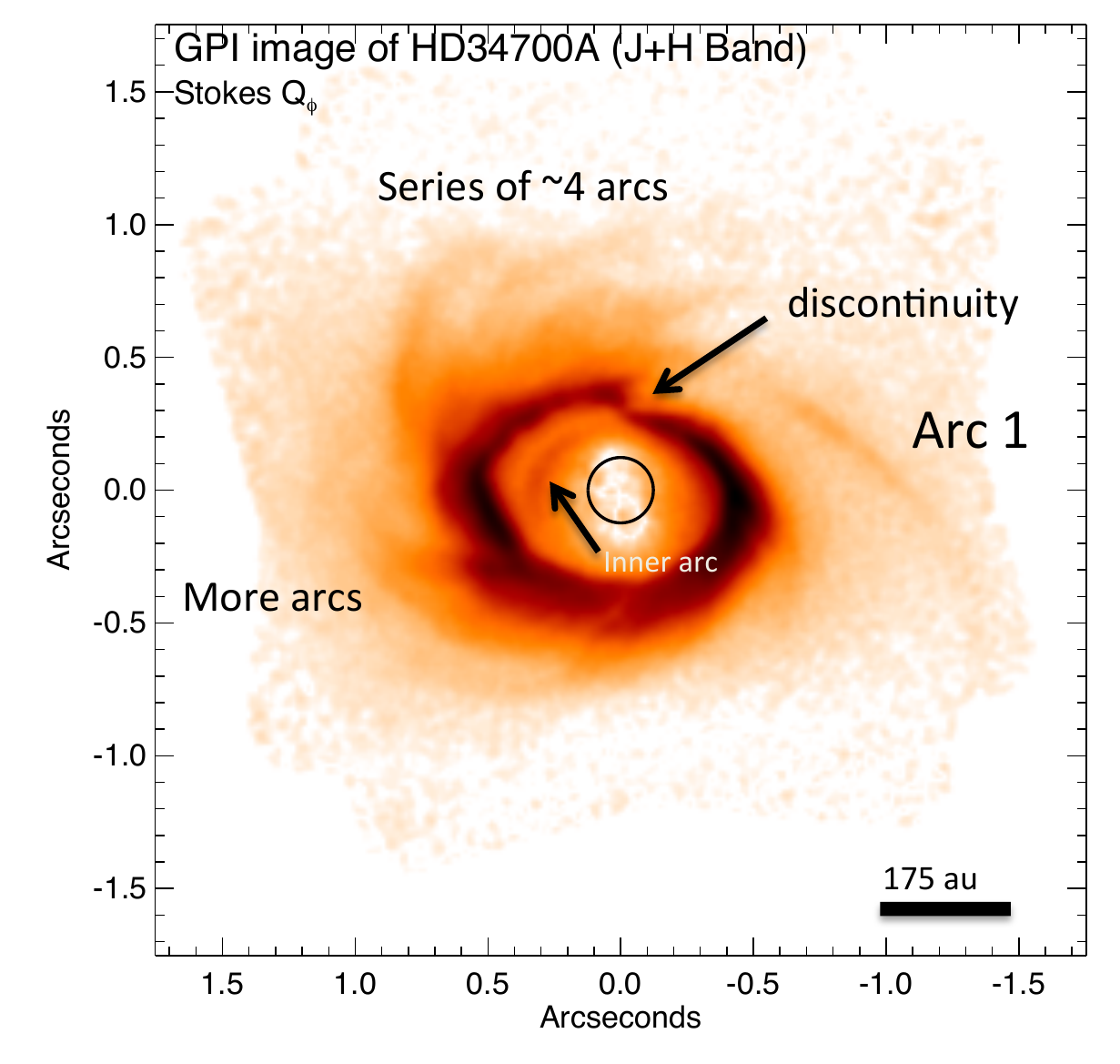}
\caption{Combined J-band and H-band $Q_\phi$ maps shown using a color table proportional to the square-root of absolute value of polarized intensity $Q_\phi$.  The maps were smoothed by a Gaussian with FWHM 30~milliarcseconds (2.1~pixels) to improve SNR and to highlight low surface brightness features outside the main ring.   The approximate location and size of the occulting spot is marked with a black circle. East is left, North is up. Features marked here are discussed in the text. }
\label{fig:description}
\end{figure}

In order to see the faint features outside the ring, we also present an image where we have combined the J and H band images based on signal-to-noise ratio and labeled major features (see Figure~\ref{fig:description}). 
 The dominant feature is an elliptical ring with major axis diameter of $\sim$1.0$\arcsec$.  The ring marks the outer edge of a lower density cavity and the beginning of  the outer dust disk, as supported later in this paper by radiative transfer modeling.  This ring has a marked discontinuity to the North, showing a sudden change in radius.  The ring is brightest in polarized intensity to the East and West, while the total intensity appears brightest to the North. 
We note this is similar to the ring brightening in $Q_\phi$ seen in the HD~163296 (MWC~275) disk recently observed by \citet{garufi2014} and \citet{monnier2017} but different from the general pattern seen in T Tauri stars where the near-side of the tilted disk is brightest in polarized light \citep{avenhaus2018}.  

Based on the brighter North side and the shift of slight off-center location of the ellipse (see next section for more detail on ellipse fitting), we expect the North side of the disk to be tilted towards us while the South side of disk tilted away from us.
While our flux calibration carries large photometric uncertainties, we report the total polarized flux at J band amounts to $\sim1.5\%$ of the total J band flux, and the polarized flux at H band amounts to $\sim2.5\%$ of the total H band flux.

 The cavity inside this ring is not devoid of scattered light and one sees an inner arc to the East that is more prominent at J band than H band.  We also marked the most extended spiral arm as ``Arc 1'' which extends from 0.5$\arcsec$ to the North out to 1.55$\arcsec$ to the West.  There is a group of roughly 4 arcs to the North-Northeast and hints of additional arc segments to the South and Southeast.  We estimate the pitch angle (angle between arc and circle tangent) of Arc 1 to be $\sim$20$\arcdeg$ at 1.5$\arcsec$ while the closer arcs to Northeast have pitch angle $\sim$30$\arcdeg$ at 0.75$\arcsec$.

 In the next section we will analyze these features more quantitatively before carrying-out radiative transfer modeling on a physical model. 

\subsection{Analysis of HD~34700A Ring and Spiral Structures}
\label{hd34700_spirals}

\begin{figure}
\centering
\includegraphics[width=4in]{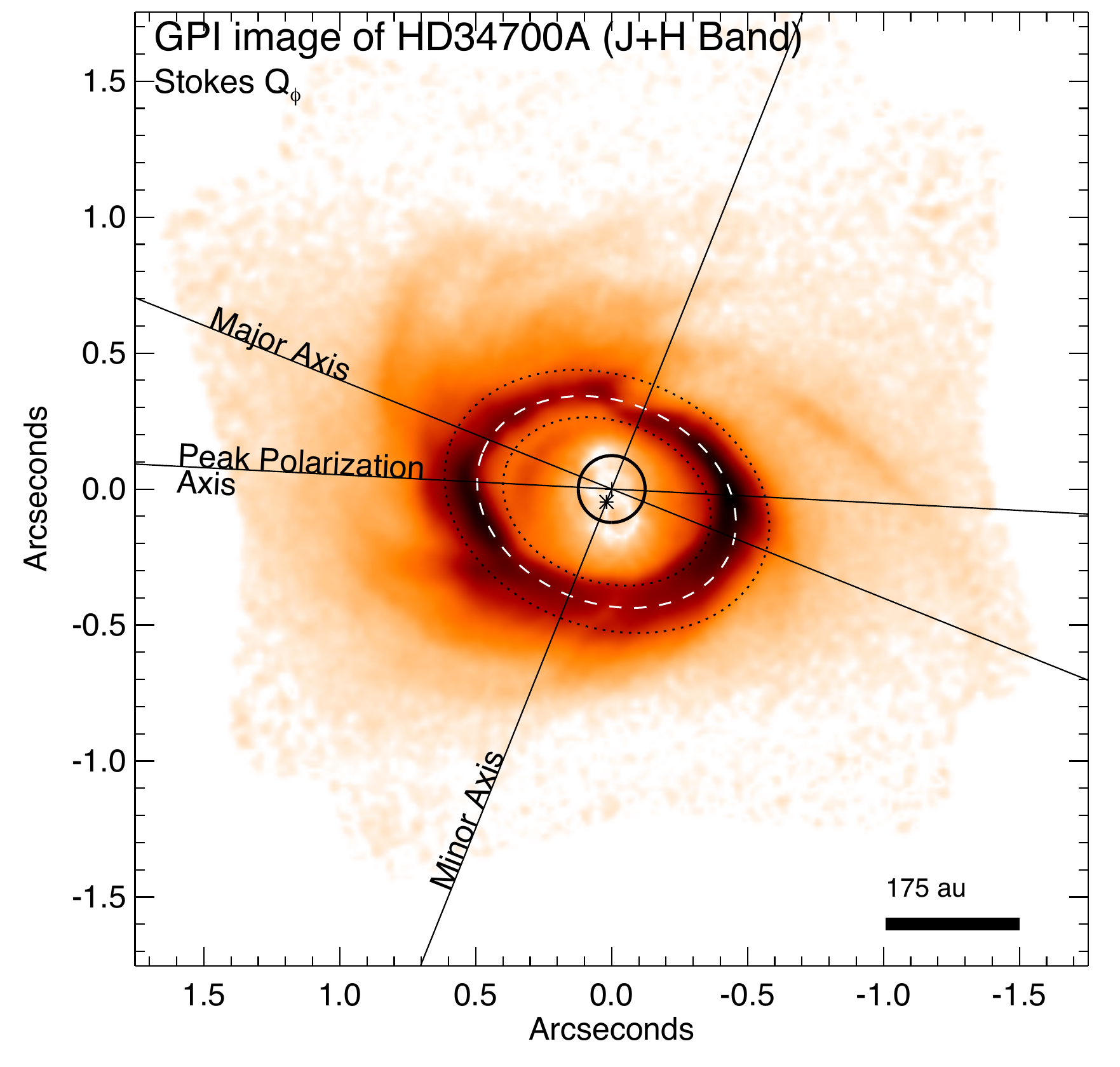}
\caption{Geometric analysis of the HD~34700A circumstellar dust ring.
We mark the axes used for radial  and annular regions used in profiles cuts in later figures.
{We fitted an ellipse to the bright ring, shown here as a dashed line.}
From this fit, we estimate a major axis of $R=0.49\arcsec=175$au inclined at 42$\arcdeg$ with elongation oriented along PA 69$\arcdeg$ East of North.  The center of this ellipse  (marked by asterisk) is shifted by 0.05$\arcsec$ South of the measured location of the central unresolved binary.  
}
\label{fig:ellipse}
\end{figure}

In Figure~\ref{fig:ellipse} we define some regions of the image for further analysis.  We first fitted an ellipse to the ridge structure around the $Q_\phi$ ring {seen in the combined J$+$H image of Figure~\ref{fig:description}.}  We constrained the center of the ring to lie along the minor axis direction, as expected for dust scattering off the surface of a flared disk and which has been seen in other disks such as MWC~275 \citep{monnier2017} and many T Tauri disks \citep{avenhaus2018}.  {The fitting procedure was as follows: 1) Starting from the peak in the Q$_\phi$ map, we followed the local maximum of the ring in both the clockwise and counter-clockwise directions.  2) These (x,y) points were fitted to an ellipse using the ``least orthogonal distance'' method as implemented in the MPFITELLIPSE routine from the IDL MPFIT library maintained by Craig Markwardt\footnote{https://www.physics.wisc.edu/~craigm/idl/fitting.html}.  We expect the uncertainty in the final parameters to be dominated by irregular structures in the ring and not the pixel-based noise, such as photon noise or variations in imaging quality.  To account for this we compiled the list of (x,y) points along the ring, pruned the list to avoid spatial correlations between neighboring points induced by the PSF, then we bootstrap-sampled these points and re-fit for an ellipse many times.  The values reported below included errors from this procedure.  }

{From this fit, we estimate a major axis of $R=0.492\arcsec\pm0.012\arcsec=175\pm5$\,au inclined at 41.5$\arcdeg\pm2.3\arcdeg$ with elongation oriented along PA 69.0$\arcdeg\pm2.3\arcdeg$ East of North.  The center of this ellipse is shifted South by 0.051$\arcsec\pm0.006\arcsec$ from the measured location of the central unresolved binary. } These features including the major and minor axis are marked in Figure~\ref{fig:ellipse} and will be used for creating radial plots.  In addition an annular region within $\pm$25\% of the best fit ellipse has been marked and will be used for azimuth profiles.

The peak of the $Q_\phi$ does not lie along the major axis of this ellipse.  Originally we thought this was due to dust density or wall height variations around this somewhat-irregular ring. However, radiative transfer modeling suggests that the peak polarization is more diagnostic of the true position angle of the inclined disk.  We mark the angle of peak polarization on this figure as well and note this angle agrees better with the position angle separating positive/negative $U_\phi$ regions in Figure~\ref{fig:ur}.

\begin{figure}
\centering
\includegraphics[width=4in]{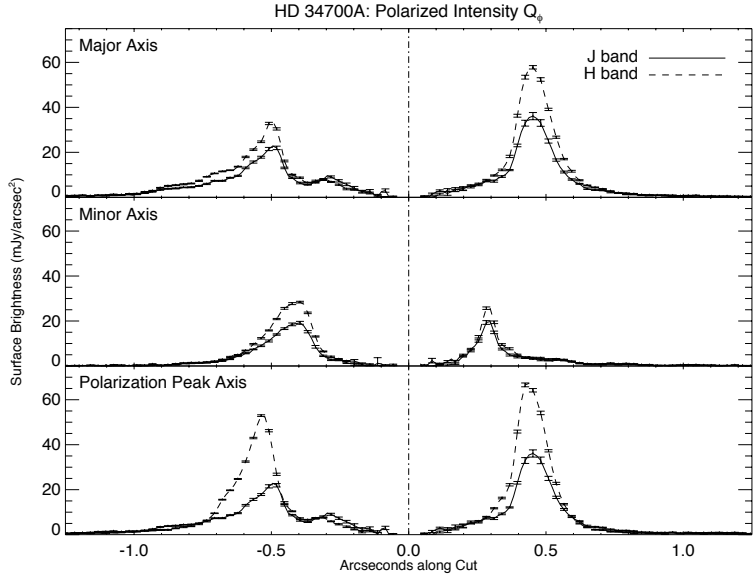}
\caption{Here we show the radial surface brightness profiles along the 3 axes defined in Figure~\ref{fig:ellipse}.  The surface brightness was averaged over a aperture 0.1'' wide. Errors bars represent the rms variation in the bootstrap sampling of the 8 independent Stokes datadubes in our observations. }
\label{fig:cuts}
\end{figure}

In Figure~\ref{fig:cuts} we show the radial profiles along the 3 axes defined in Figure~\ref{fig:ellipse} of the $Q_\phi$ surface brightness at J and H bands.  One can see the flux from the``inner arc'' appearing at 0.3$\arcsec$ along the major axis, showing relatively larger flux levels ($>$50\%) at J band than H band compared to the rest of the ring.  The errors were calculated by bootstrap sampling of the 8 independent Stokes data cubes.

\begin{figure}
\centering
\includegraphics[width=6.in]{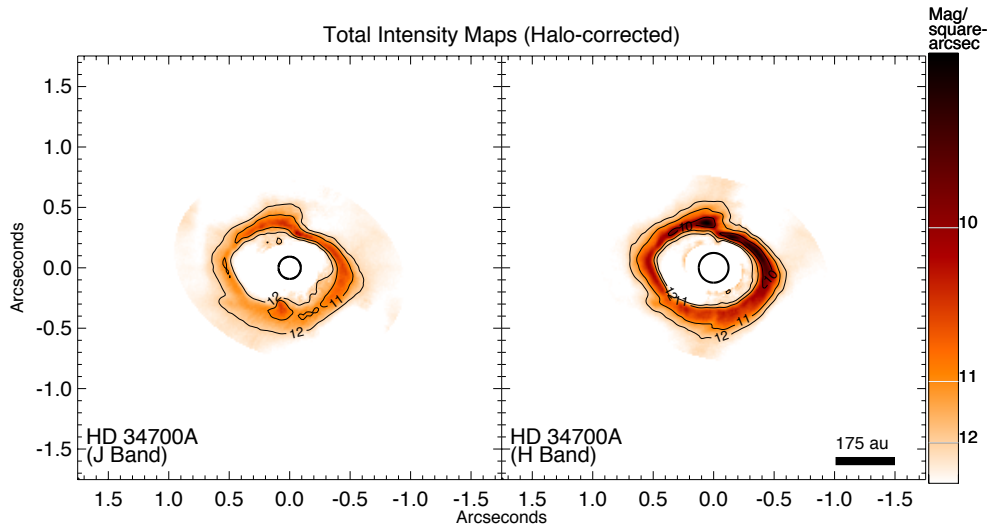}
\caption{This figure shows the total intensity of the scattered light around HD~34700A, extracted by subtracting a Moffat PSF model for the halo surrounding the coronagraphic spot.  The surface brightness is shown with a linear color table. The subtraction is only valid within $\pm$30\% of the ring itself. \label{fig:toti_halo}}
\end{figure}

{Next we want to analyze the intensity around the dominant ring, identified as the inner wall of the dusty transition disk.  For this particular source, we were able to extract a crude estimate of the actual total intensity of the circumstellar dust scattering by subtracting a model of the instrument point spread function, something not normally possible to do with GPI data.  Specifically, a Moffat function
was used to approximate the PSF by fitting the total intensity image with the annular ring masked out.  We used the MPFIT2DPEAK function in the IDL MPFIT library (also maintained by Craig Markwardt).  Defining our Moffat function as $I(x,y) = A_0 + {A_1} {(u +1)^{-A_7}}$ where $u = ( (x^\prime-A_4)/A_2 )^2 + ( (y^\prime-A_5)/A_3)^2)$ and $(x^\prime,y^\prime)$ is a reference frame tilted by angle $A_6$). We restricted the fitting region to just within two annuli positioned inside and outside the best-fit ellipse that defines the ring (specifically, region 1 was between 0.4-0.7$\times$ the ring and region 2 was between 1.35-2.5$\times$ the ellipse). This allows us to approximate the power-law point source function and extract the extra scattered light coming from the ring region clearly seen in $Q_\phi$ -- see Figure~\ref{fig:toti_halo}.  For completeness we include the Moffat function parameters we adopted for the J and H band total intensity extraction, along with errors based on the bootstrap sampling of the 8 independent Stokes datacubes: 
$A_J = (-1.7\times10^{-5}\pm0.2\times10^{-5}$,
$0.019\pm0.005$,
$0.010\arcsec\pm0.003\arcsec$,
$0.008\arcsec\pm0.002\arcsec$,
$0.014\arcsec\pm0.018\arcsec$,
$0.012\arcsec\pm0.018\arcsec$,
$ 37.5\arcdeg\pm4.0\arcdeg$ E of N,
$ 0.654\pm0.020 )$ ; 
$A_H = ( -2.5\times10^{-5}\pm0.4\times10^{-5}$,
$0.0036\pm0.0007$,
$0.021\arcsec\pm0.005\arcsec$,
$0.016\arcsec\pm0.004\arcsec$,
$-0.013\arcsec\pm0.002\arcsec$,
$-0.007\arcsec\pm0.002\arcsec$,
$ 50.9\arcdeg\pm1.3\arcdeg$ E of N,
$ 0.526\pm0.024 )$.   
}

With the estimate of the scattered light total intensity {shown in Figure~\ref{fig:toti_halo}}, we can calculate the true fractional polarization not just $Q_\phi$. As part of this analysis, we searched for point sources within the halo and noted two symmetrical spots in the J band image at radius 0.3'' and position angle -12$\arcdeg$/168$\arcdeg$ (slightly evident in Figure~\ref{fig:toti_halo}).  These spots appear right on the main ring but do not appear in the H band total intensity images.  Given the symmetry of the spots and the lack of H band detection, we identify these features as adaptive optics artifacts, i.e., not physical companions such as exoplanets.

\begin{figure}
\centering
\includegraphics[width=4in]{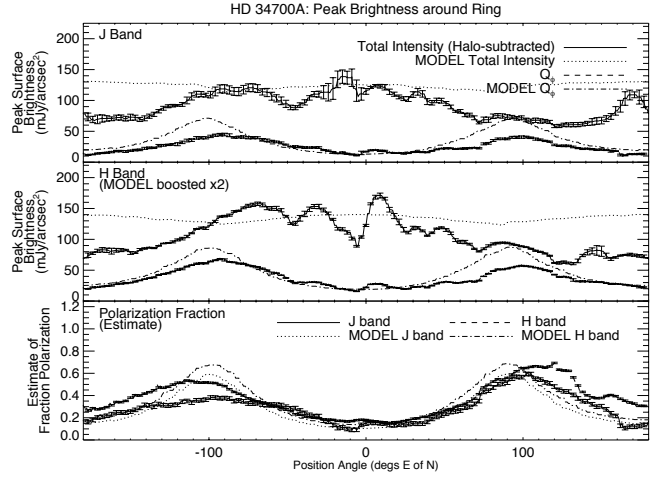}
\caption{This figure shows the total scattering intensity along with the Q$_\phi$ intensity as a function of azimuth around the strong circumstellar dust ring.  (top) J band results, (middle) H band results, (bottom) Estimate of the fractional polarization.  We see the most scattering comes from the North, identified as the near side of the disk, while the highest fractional polarization is on the East/West sides of the ellipse.  Results from our radiative transfer model are also presented and reproduce the fraction of polarization and azimuthal variations of $Q_\phi$.  {The H-band model intensities are systematically lower than our observed values and we have multiplied the model H band intensities by 2 to ease comparison with our H band flux levels. }}
\label{fig:ann_cuts_qr}
\end{figure}

\begin{figure}
\centering
\includegraphics[width=4in]{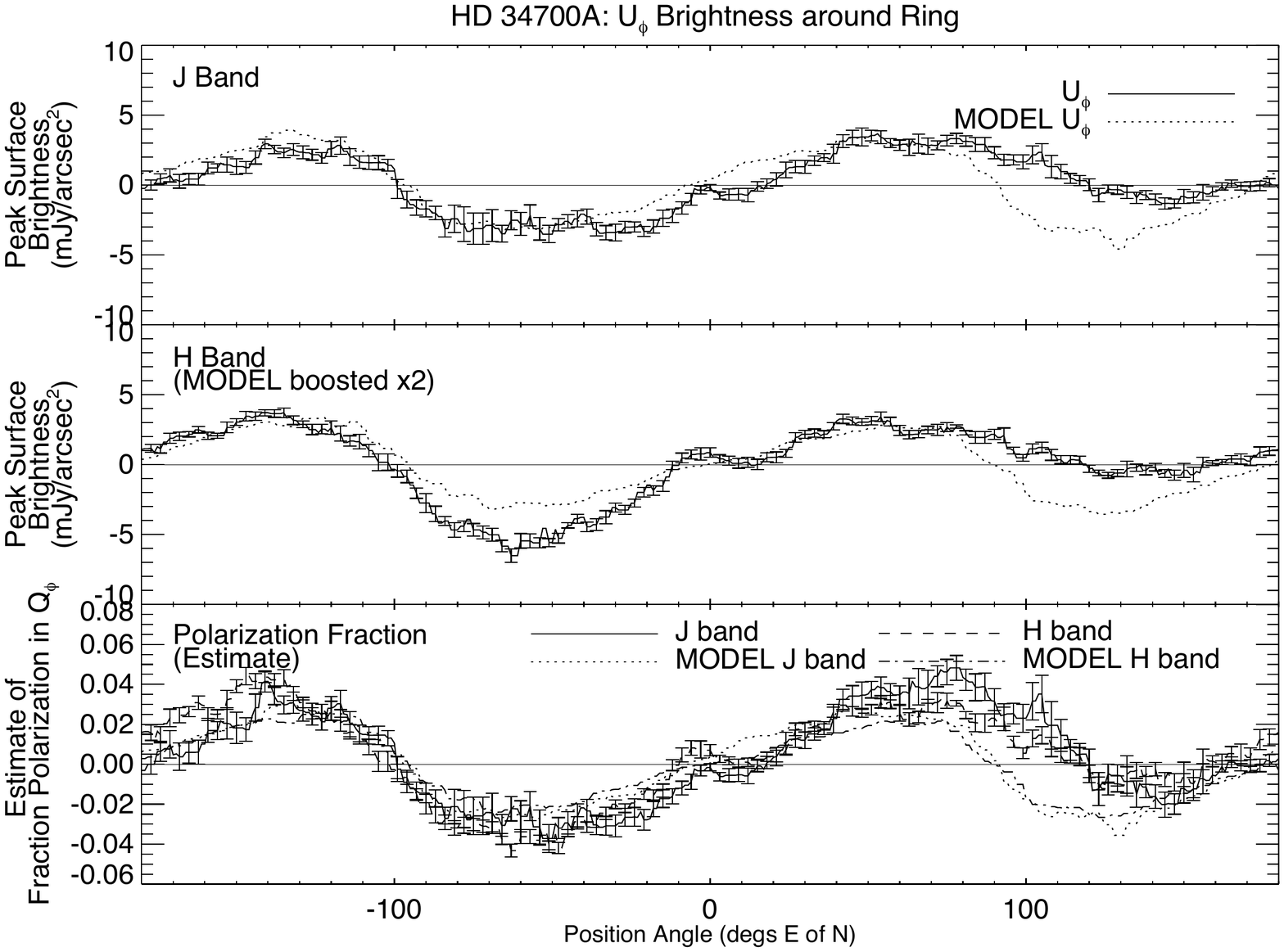}
\caption{This figure shows the total scattering intensity along with the U$_\phi$ intensity as a function of azimuth around the strong circumstellar dust ring.  (top) J band results, (middle) H band results, (bottom) Estimate of the fractional polarization.  We see strong $U_\phi$ modulation as a function of position angle. The radiative transfer model presented in \S\ref{model} reproduces this feature and we have included the model results in the figure as well. {The H-band model intensities are systematically lower than our observed values and we have multiplied the model H band intensities by 2 to ease comparison with our H band flux levels. }  }
\label{fig:ann_cuts_ur}
\end{figure}

Armed with the halo-subtracted total intensity of the scattered-light disk, we constructed Figure~\ref{fig:ann_cuts_qr} where the peak $Q_\phi$ surface brightness around the ring is plotted as a function of position angle within the annulus defined in Figure~\ref{fig:ellipse}.  Note we used the $Q_\phi$ peak locations and found the corresponding total intensity at that location for the other observables found in Figure~\ref{fig:ann_cuts_qr}.  
We see the peak total intensity varies by about 50\% around the ring while the $Q_\phi$ varies by nearly a factor of 3.  
The absolute fractional polarization peaks at about 50\% for J band and about 60\% at H band.  Figure~\ref{fig:ann_cuts_ur} contains similar profiles around the ellipse for $U_\phi$, also associated with locations where $Q_\phi$ is maximum at each position angle.  We see the fractional $U_\phi$ polarization varies $\pm$3\% around the ring.

As part of our analysis of HD~34700A, we analyzed archival Hubble Space Telescope imagery from 1998.  The results are interesting though preliminary and we have included details in Appendix~\ref{appendix_hst}.  In short, we find some evidence that the scattered light ring has rotated 5.75$\pm$0.25\arcdeg (error bar not including possibly large systematic errors) counter-clockwise over the 19.3 years since the HST data was taken -- the spiral arms winding is consistent with counter-clockwise motion.  This rotation implies an orbital period of 1200$\pm$50\,yr, consistent with the expected 1160\,yr period for material at 175au around the central binary with a combined mass of 4\,M$_\odot$.  We also tentatively identified a candidate brown dwarf at 6.45'' distance (projected 2300au), but confirmation of common proper motion has not been made yet. If confirmed, HD~34700ABCD would be a rare young quintuplet system that seems highly unstable from a dynamical point of view.

We will discuss possible origins of the marked discontinuity on the North side of the ring in later sections, including the possibility of shadowing affects from an inner disk (see \S\ref{model}) or from material around an vigorously accreting protoplanet (see \S\ref{hydro}).

Next, we develop a radiative transfer model and will compare the model results to the observed profiles we just discussed.

\section{Radiative transfer modeling of the HD~34700A system}
\label{model}

Here we focus on creating a physical model for HD~34700A that can simultaneously explain the bright polarized ring and spectral energy distribution using a self-consistent two-dimensional radiative transfer calculation; future work will address the complex spiral structures and possible shadowing effects when additional data are available. {Since we know {\em a priori} that our model will not fit the data well in detail, we chose to manually adjust our model parameters to achieve a qualitatively good fit to both the imaging and the SED, whilst constraining as many free parameters as possible to their canonical values. We estimate the errors on each of our derived parameters by calculating the range of models (fixing the other parameter) that give an equally good quantitative fit.}

The modeling was conducted using the   {\sc torus} radiative transfer code \citep{harries_2000,harries_2004,harries_2011} which uses the Monte Carlo (MC) radiative equilibrium method of \citet{lucy_1999} implemented on an adaptive mesh. The {\sc torus} code has been extensively benchmarked against analytical solutions and 1-D and 2-D test problems \citep{harries_2004, pinte_2009}. 

The photometric data compilation of \cite{seok2015} was adopted in this study {(see their Table~1 for details)}. This comprises {\it EXPORT} $UBVRI$ photometry \citep{mendigutia2012} plus 2MASS, {\it WISE}, {\it AKARI}, {\it IRAS} and {\it SCUBA} data culled from online catalogs. 

\subsection{Stellar parameters}
\label{stars}
{
The new {\em Gaia} distance of 356.5~pc is nearly 3$\times$ farther away than expected by the spectroscopic analysis of \citet{torres2004} and demands a re-evaluation. We have updated the masses and ages for HD~34700A binary as follows: 1) We start with $T_{\rm eff}=$5900K, 5800K and V mag $=$9.85, 9.95 \citep[directly from][]{torres2004}. 2) We convert apparent magnitude to absolute magnitude using the new distance, yielding $M_V=$2.09, 2.20.  3) Assuming solar metallicity and using the pre-main sequence tracks of \citet{siess2000}\footnote{Using web interface at http://www.astro.ulb.ac.be/~siess/pmwiki/pmwiki.php?n=WWWTools.HRDfind}, we find HD~34700A to be a young system ($\sim5$\,Myrs) consisting of two 
$\sim$2.05\,M$_\odot$ stars with a transition disk.}

{Fixing the stellar temperatures and luminosity ratio ($l_2/l_1=0.9$) based on the arguments above, we can determine the stellar radii by  fitting Kurucz model atmospheres  to the optical/near-IR photometry, fixing the distance at 356\,pc and allowing the extinction ($A_V$) to vary while fixing the total-to-selective extinction to its canonical Milky Way value ($R=3.1$).}

{
A fit using $UBVRI$, $JHK$ and WISE 3.4\,$\mu$m and 4.6\,$\mu$m fluxes gives $R_1$=3.80\,R$_\odot$, $R_2$=3.73\,$_\odot$ and $A_V=0.2$ although the fit is rather poor (reduced chi-squared $\overline{\chi^2}=166$). Restricting the photometry to just $UBVRI$ and $JHK$ gives a vastly improved fit ($\overline{\chi^2}=18$) with $R_1=3.46$\,R$_\odot$, $R_2=3.40$\,R$_\odot$ and $A_V=0$, strongly indicative of a disk excess longwards of 3.4\,$\mu$m. We therefore adopted stellar radii of $R_1=3.46$\,R$_\odot$, $R_2=3.40$\,R$_\odot$ and zero reddening for our radiative transfer models. 
}

\subsection{The disk structure}
\label{sec:diskstructure}

The disk density in cylindrical coordinates $(r,z)$ is given by
\begin{equation}
\rho(r,z) = \rho_{\rm m} \left( \frac{R_{\rm i}}{r} \right)^{\alpha} \exp \left( -\frac{z^2}{2 h(r)^2} \right)
\label{density_eq}
\end{equation}
where $\rho_{\rm m}$ is a the midplane density at the inside edge of the transition disk at radius $R_{\rm i}$, $\alpha$ is the density power-law index, and the scale-height $h(r)$ is given by 
\begin{equation}
h(r) = h_{\rm i} (r / R_{\rm i})^\beta
\label{scaleheight_eq}
\end{equation}
where $h_{\rm i}$ is the scale-height at $R_{\rm i}$ and $\beta$ is the flaring index. The value of $\rho_{\rm m}$ is found from
\begin{equation}
M_{\rm disk} = \int_{R_{\rm i}}^{R_{\rm o}} \int_{-\infty}^{\infty} 2\pi r\rho(r,z)\,dz\,dr 
\end{equation}
where $R_{\rm o}$ is outer disk radius. We assume that $R_{\rm o}$ marks a sharp cutoff to the outer disc.

\subsection{Modelling procedure}

{The disk structure was discretised on an adaptive cylindrical mesh, initially defined so that the disk was sampled vertically such that there were at least three cells per scale-height at all radii. This initial grid was then further adaptively refined in order that sharp opacity gradients (such as the disk inner edge) were adequately resolved, a step which is essential to capture the temperature gradients and produce the correct SED (particularly in the near-to-mid infrared). This was achieved by iteratively refining those optically thick cells ($\tau > 1$) that neighboured optically thin cells ($\tau < 0.1$). This resulted in a mesh composed of approximately 250,000 cells.}

{The radiative equilibrium procedure started with a 1 million photon packets for the first iteration, with the number of photon packets doubling at each iteration until the emissivity of the dust integrated across the entire volume converges to a tolerance of 1 per cent (indicating the temperatures are well converged). This typically takes 5 iterations. The SEDs and images are then computed for a particular inclination using 10 million photon packets.}

\subsection{Modeling strategy and the best fit model}

The outer disk radius ($R_o$) is poorly constrained by SED fitting. However we note that ``Arc 1'' marked in Figure~\ref{fig:description} extends to a radius of 1.5\arcsec, corresponding to a linear radius of $\sim 500$\,au, so we adopted this value as our outer disk radius. The inside edge of the transition disk is defined as the radius of the bright ring (0.5\arcsec or 175\,au) as measured directly from the polarization images. We also fixed the inclination and position angle of the disk from the ellipse fitting results (see \S\ref{hd34700_spirals}).

The brightening of the ring to the East and West in polarized light (see Figures~\ref{fig:qr} and~\ref{fig:ann_cuts_qr}) is due to scattering by  grains that have polarizability that peaks at scattering angles close to~$90^\circ$. Although it is possible to achieve this type of polarization behavior with dust distributions that incorporate larger ($>1$\,\micron) grains, the forward-scattering nature of such dust leads to a significant excess of polarized light in the front part of the ring, and a corresponding deficit to the rear. We therefore include a component of small (0.1\,\micron) silicate grains in the model, that effectively scatter in the Rayleigh limit. However we find that a contribution of larger grains is necessary to simultaneously fit the SED and the imaging, and we include a second distribution of dust with sizes between 1--1000\,\micron\ with an MRN power-law distribution \citep{mathis1977}. {We assume a dust ratio of 50:50 small grains to large grains by mass, since this is not particularly well constrained by the modelling. However the {\em total} mass of dust is quite well constrained by the long wavelength part of the SED.}

It can clearly be seen from Figure~\ref{fig:ellipse} that the maximum polarization axis is offset from the major axis of the ellipse fit. Such an offset might occur if the bright inner edge of the disk was not truly circular, for example if it was formed from a pair of tightly-wound spiral arms. We therefore investigated whether the polarization variation around the ring could give a more reliable measure of the inclination and position angle of the disk inner edge. If we assume the Rayleigh scattering phase matrix is applicable, and assume a thin ring illuminated by an unpolarized central source that scatters just once, the polarized flux $P$ around the ring should vary as
\begin{equation}
P(\theta) = \gamma \sin^2 \theta
\end{equation}
where $\theta$ is the scattering angle and $\gamma$ is a constant scaling factor that depends on the intensity of the source. The angle $\theta$ is related to the azimuthal angle ($\phi$) around the ring as
\begin{equation}
\cos(\theta) = \cos(\phi-\delta) \sin i
\textit{}\end{equation}
where $i$ is the inclination and $\delta$ is the position angle of minimum polarization. We fitted the above equations to the J-band $Q_\phi$ curve given in Figure~\ref{fig:ann_cuts_qr} {using a $\chi^2$ grid search, and the best fit ($\overline{\chi^2}=3.3$)} was found with $i=43^\circ$, $\gamma=51$, and $\delta = 4\arcdeg$. Hence the geometry of the ring determined from the polarization distribution gives a similar inclination to the elliptical fit, but a position angle of the major axis that is significantly closer to East-West

{We fixed the power-law flaring index of the disk ($\beta$) to its canonical value of 1.125 \citep{kenyon1987} and the radial density power law index ($\alpha$) to $-2.125$ which fixes the radial surface density power-law index to $-1$.} A more flared disk (i.e. a higher $\beta$) gives a polarized surface-brightness profile that is marginally shallower than the observations. With other parameters fixed, the peak of the IR thermal emission in the SED is controlled by the scale-height of the inner disc. We find a value of {$17\pm 2$}\,au matches the SED, giving an $h/r$ of the disk at 175\,au of 0.1.

It has previously been established that a contribution from very small grains (VSGs) and polycyclic aromatic hydrocarbons (PAHs) is required to fit the mid-IR spectrum of HD~34700A \citep{seok2015}. We have implemented the microphysics associated with VSG/PAH emission according to the prescription of \cite{robitaille2012}, which in turn is a modification of the method of \cite{wood2008}. For this model we kept the same silicate dust composition as the previous best fit solution, but reduced the scale height of the inner edge {to $9 \pm 2$\,au} in order to compensate for the additional IR emission from VSGs. The final parameters of the model presented here can be found in Table~\ref{params_tab}. We find reasonable agreement with the {\it Spitzer} IRS spectrum, but we are still underestimating the flux at 3.6 and 4.5\,\micron. The \citet{seok2015} SED fit showed a similar deficit, although the stellar parameters they used meant that they over-estimated the near-IR and thus their model mid-IR flux was higher than ours.

Note that we make no strong claims about the uniqueness of our model parameters, particularly for those whose leverage derives primarily from the SED. In fact previous modellers have demonstrated adequate SED fits when the object was thought to be a Vega-like debris disk system at 55\,pc with a 50\,au cavity and a geometrically- and optically-thin power-law density distribution \citep{sylvester1996}. Fortunately the {\it Gaia} DR2 distance has settled the largest ambiguity in the modeling, indicating that the system is a pre-main-sequence binary with a transitional disk and providing a much tighter constraint on the total dust mass.  In combination with the distance we also possess new constraints from our imaging data, in particular the location of the disk inner edge and the surface brightness profile in scattered light (which constrains the disk flaring). The scale height of the inner edge is then determined by both the mid-IR peak of the SED, and by the polarized surface brightness of the imaging (see section~\ref{model_comp_sec}).

\begin{table}
\begin{center}
\caption{Model parameters for HD34700A}
\label{params_tab}
\begin{tabular}{lll}
\hline
Parameter & Value & Description \\
\hline
\multicolumn{3}{c}{Stellar parameters} \\
{Primary stellar radius, $R_1$} &  3.46 \,R$_\odot$  & Fitted from photometry.\\
{Secondary stellar radius, $R_2$} &  3.40 \,R$_\odot$  & Linked to $R_2$ via luminosity ratio.\\
{Primary effective temperature} & 5900\,K & Fixed; \citet{torres2004}. \\
{Secondary effective temperature} & 5800\,K & Fixed; \citet{torres2004}. \\
Stellar masses, $M_*$ & 2\,M$_\odot$ & \citet{torres2004}.\\
Distance                  & 356\,pc & \cite{gaia2018}.\\
$A_V$ & 0 & Fitted from photometry. \\
\hline
\multicolumn{3}{c}{Orientation parameters} \\
Inclination, $i$ & 43$^\circ$ & Fixed from ellipse fit. \\
PA of max. polarization& 86$^\circ$ & Fixed from polarization fit. \\
\hline
\multicolumn{3}{c}{Disk parameters} \\
Disk dust mass, $M_{\rm disk}$ & {$1.2 (\pm 0.2) \times 10^{-4}$\,M$_\odot$} & Fitted via SED.\\
Disk flaring index, $\beta$ & 1.125 & Fixed at canonical value.\\
Radial density index, $\alpha$ & $-2.125$ & Fixed at canonical value.\\
Inner disk radius, $R_{\rm i}$ & 175 \,au & Fixed from imaging measurement.\\
Scale-height at inner radius, $h_i$ & {$17 \pm 2$ \,au} & No PAH/VSG model. Fitted from SED. \\
Scale-height at inner radius, $h_i$ & {$9 \pm 2$ \,au} & PAH/VSG model. Fitted from SED. \\
Outer disk radius, $R_{\rm o}$ & 500\,au & Fixed from image measurement.\\
\hline
\multicolumn{3}{c}{Grain properties, small grains} \\
Grain type    & Silicates & \citet{draine_1984}. \\
Grain size, $a_{\rm small}$ & 0.1\,$\mu$m & Fixed. Required from imaging. \\
Fraction of dust mass   & $0.5$ & Fixed.\\
\hline
\multicolumn{3}{c}{Grain properties, large grains} \\
Grain type    & Silicates & \citet{draine_1984}. \\
Min grain size, $a_{\rm min}$ & 1\,$\mu$m & Fixed. Required to fit SED.\\
Max grain size, $a_{\rm max}$ & 1000 \,$\mu$m & Fixed. Required to fit  SED. \\
Grain size power law index & $-3.5$ & Fixed; \citet{mathis1977} \\
Fraction of dust mass  & $0.5$  & Fixed. \\
\hline
\end{tabular}
\end{center}
\end{table}

\begin{figure}
\centering
\includegraphics[width=6in]{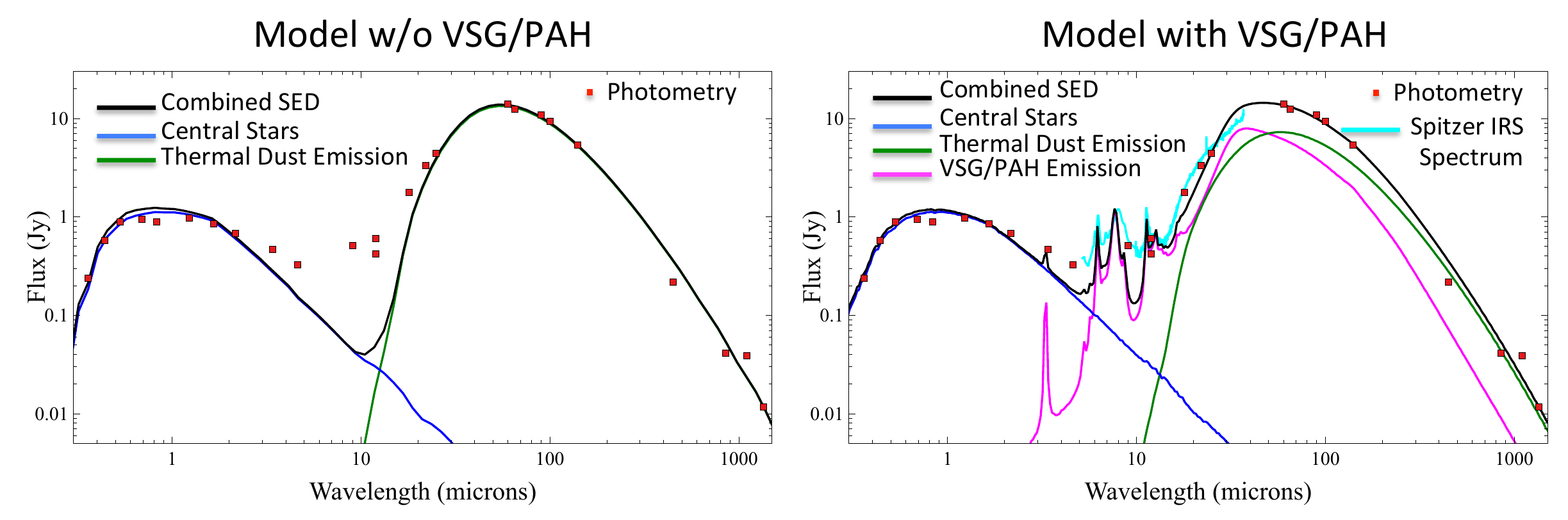}
\caption{Radiative-transfer models of HD~34700A compared with photometric observations. The left-hand panel shows the model without VSG/PAH emission and includes the direct photospheric radiation from the central star (blue line), the thermal dust emission (green line), the total SED (black line), the photometric points (red squares). The model that includes VSG/PAH emission is shown in the right-hand panel. The meaning of the colors is the same as the left hand-panel, and the contribution to the SED of VSG/PAH emission is shown (pink line) along with the {\it Spitzer} IRS spectrum (cyan line).}
\label{fig:sed_fit}
\end{figure}

\subsection{Comparison of Models to Data}
\label{model_comp_sec}
\begin{figure}
\centering
\includegraphics[width=6in]{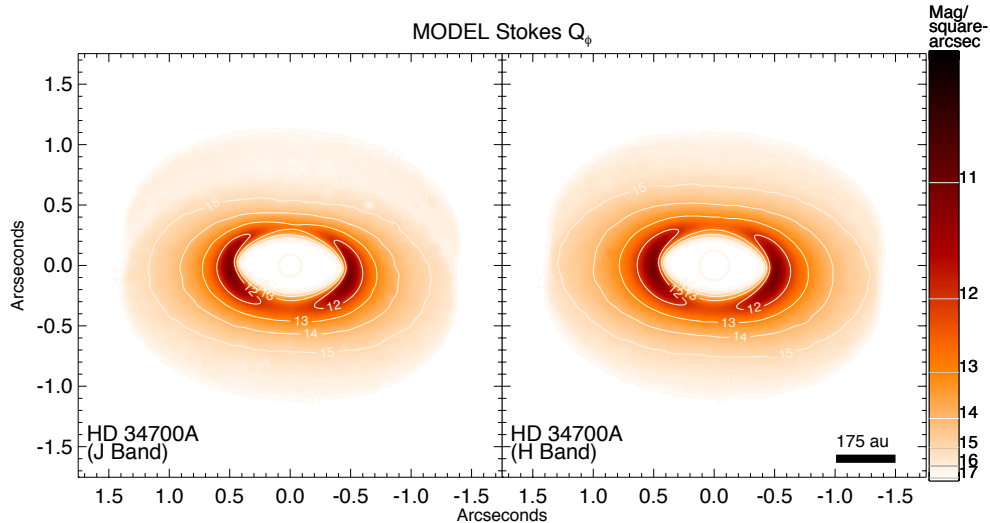}
\caption{Simulated $Q_\phi$ images of HD~34700A in J-band and H-band based on the radiative transfer model w/o VSG/PAH described in the text, smoothed to the same angular resolution as our data.  The model captures the basic shape and location of the ring, including the East-West brightening. }
\label{fig:model_qr}
\end{figure}

\begin{figure}
\centering
\includegraphics[width=6in]{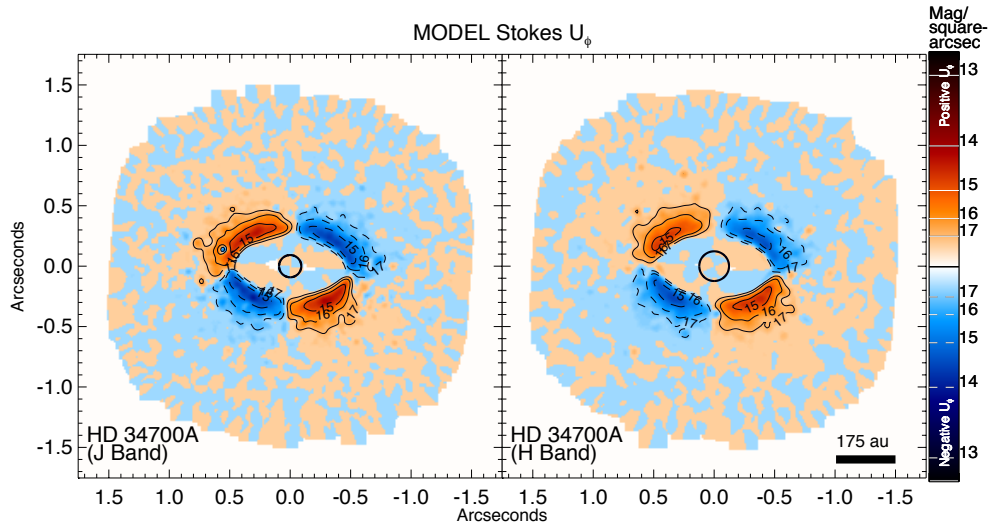}
\caption{Simulated $U_\phi$ images of HD~34700A in J-band and H-band based on the radiative transfer model w/o VSG/PAH described in the text, smoothed to the same angular resolution as our data.  The radiative transfer modeling naturally produces the ``butterfly'' pattern seen in our data (see Figure\,\ref{fig:ur}), {supporting the claim that the observed $U_\phi$ to be from multiple scatterings in an optically thick disk, and not due to miscalibrations or systematic errors in the pipeline.}
}
\label{fig:model_ur}
\end{figure}

Figure~\ref{fig:sed_fit} shows the results of the model SED compared to the collected photometry. We see that VSG/PAH emission is needed to explain the emission between 3-20$\mu$m.  {For comparing to GPI data, we only show results for the model w/o VSG/PAH for simplicity, since both models give similar results for J and H band. }  Figure~\ref{fig:model_qr} shows the azimuthal-component of the polarized intensity $Q_\phi$ for the model w/o VSG/PAH, reproducing the strong 
East-West brightening along the main ring.   More impressive might be Figure~\ref{fig:model_ur} which shows the model $U_\phi$ surface brightness has the same butterfly pattern seen in the real $U_\phi$ data in Figure~\ref{fig:ur} -- confirming the pattern predicted by \citet{canovas2015}.  

We can be more quantitative in our comparison of the model and GPI images -- we have included the model intensities for $Q_\phi$ and $U_\phi$ in Figures~\ref{fig:ann_cuts_qr} and \ref{fig:ann_cuts_ur}.  Here we see excellent agreement at J band (where the model was optimized) with good $Q_\phi$ and $U_\phi$ agreement, although the total intensity varies less azimuthally in these models than in our data.  The overall model H band intensities are 2x smaller than the observed fluxes and we have boosted them in these figures to make comparison to data easier.  Note that our photometric calibration of both J and H bands are poor and so we can not rule out calibration errors in explaining this x3 discrepancy.   The fractional polarization for both J and H bands does not depend on absolute photometric calibration and we see that the model polarization fraction is a bit higher than we observe3.  

More exploration of the disk parameters should improve agreement between models and data.  Qualitatively, we suggest that higher model H flux, more forward scattering on the front side of disk, and lower fractional polarization could be achieved with larger grains or varying dust constituents, e.g., ice mantles, carbon grains.

Another weakness of our current model is the adoption of a dust-free  cavity which of course can not produce the scattered light in the low-density cavity seen in our images, although one must account for the wings of the telescope/adaptive optic system PSF to determine the true intrinsic polarized flux just inside the rim.

In fact the reduced chi-squared values of  our best-fit stellar atmosphere indicate that the optical/near-IR photometric data are formally inconsistent with pure stellar emission, even when fitting the $UBVRI$ data alone. Furthermore our RT models undershoot the {\it WISE} photometry (see Figure~\ref{fig:sed_fit}), even when PAH/VSG emission is included. (We note that the fits by \cite{seok2015} demonstrate a similar behavior.)

If one were to adopt a marginally lower stellar luminosity (thus degrading the fit to the optical photometry) one could conclude there was an near-IR excess, and thus evidence for an unresolved warm ($\sim 1500$\,K) disk component very close to central binary. This putative inner disk would then cast shadows on the outer disk (an observation which in itself means that any inner disk cannot be too flared, or the ring at 175\,AU would be much fainter). Of course the inner disk might not necessarily be aligned with the outer disc, and the misalignment could contrive to produce the inner arc (see Figure~\ref{fig:description}) for example. However a misaligned inner disk should cast diametrically opposed shadows on the bright ring (e.g. \citealt{benisty2017}), which are not observed.
We defer a three-dimensional RT study of this system to a future paper.

\section{Possible Interpretation of Disk Features and Hydrodynamic Models}
\label{hydro}
While our radiative modeling allows us to explain the general appearance of HD~34700A in terms of the dust density and temperature distributions, the specific features like the spiral arms and the discontinuities seen in Figure\,\ref{fig:description} require a hydrodynamics approach.
Indeed, most of the features seen in the HD~34700A transition disk must have their origin in disk hydrodynamics. We briefly discuss relevant processes:

\begin{itemize}
\item{{\it Binary system at the center (HD~34700Aa,Ab).}
It has been recently suggested that the observed structures in the HD~142527 disk, including inner cavity and spirals \citep{avenhaus2017}, can be explained by the interaction between the disk and the binary companion at the center of the system \citep{price18}. 
The semi-major axis of the binary assumed in the models of \citet{price18} is a significant fraction of the cavity size ($26.5 - 51.3$ au vs. $\sim100$~au). 
Other numerical simulations of circumbinary disks also show that the size of the inner cavity opened by the central binary is a factor of a few larger than the binary semi-major axis \citep[e.g.,][]{pierens18}.
However, in case of HD~34700A the cavity size ($\sim175$~au) is orders of magnitude larger than the semi-major axis of the central binary (0.69~au), making this possibility unlikely to be the main origin of the cavity and spirals.
}

\item{{\it Unseen planetary companion(s) within the inner cavity.}
Having a sufficiently massive planetary companion in the cavity will naturally explain the cavity.
Regarding spiral arms, two-dimensional hydrodynamic simulations show that companion with a circular orbit can launch only one or two spiral arms exterior to its orbit \citep{baezhu18b},  because the constructive interference among wave modes to form spiral arms becomes unavailable far from the companion \citep{baezhu18a}. 
It is therefore unlikely that a single companion having a circular orbit excites all the observed spiral arms. 
When a companion has an eccentric orbit, however, it will introduce additional families of wave modes having different orbital frequency \citep{goldreich80}.
These waves can constructively interfere with each other, forming a larger number of spiral arms than a companion with a circular orbit. 
If the set of spiral arms in the Northeastern side of the disk (noted with ``series of $\sim4$ arms'' in Figure \ref{fig:description}) are driven by one unseen planetary companion within the inner cavity, for instance, it is very likely that the object has an eccentric orbit.
Alternatively, multiple planetary companions within the inner cavity can be the cause of the large number of spiral arms in the disk.
}

\item{{\it External companion (HD~34700B).}
It is possible that the external companion HD~34700B (excites spiral arms in the disk around the central binary. 
Numerical simulations showed that a companion star in a binary system can excite spiral arms in the disk around the primary star \citep[e.g., HD~100453;][]{dong16}. 
However, in case of a stellar companion whose mass is a significant fraction of the primary star (in this case primary binary), the companion generates nearly axisymmetric, $m=2$ spiral arms \citep{fung15,baezhu18b}. 
The external companion is therefore not sufficient to explain all the observed spiral arms.
{With the {\em Gaia} DR2, the distances to HD~34700A and HD~34700B are the same within parallax errors, consistent with HD~34700B being gravitationally bound with HD~34700A. Since the true three-dimensional distance between the systems is likely larger than the projected distance of 1850\,au,  it is unlikely that HD~34700B is responsible for exciting the spiral arms.  The same holds true for HD~34700C or HD~34700D which are at an even larger projected distance and without {\em Gaia} parallax yet to confirm physical association. }
}

\item{ {\it Gravitational instability.}
Based on the disk surface density and temperature profiles obtained with the radiative transfer modeling presented in Section \ref{model}, the Toomre $Q$ parameter is greater than 25 everywhere in the disk with 100:1 gas-to-dust mass ratio. The disk is therefore unlikely to be gravitationally unstable currently.
}

\end{itemize}

In order to examine the potential origin of the disk features, we carried out three-dimensional hydrodynamic simulations.
In particular, we examined whether an unseen planetary companion in the inner cavity could be responsible for the observed disk structures
using FARGO 3D \citep{benitez16,masset00} to simulate disk hydrodynamics.
The hydrodynamic simulation domain covered 54 to 810~au (0.15'' to 2.25'') in the radial direction, 15$^\circ$ above and below the midplane in the meridional direction, and the entire $2\pi$ in the azimuthal direction.
We used the disk density and temperature profiles described in Section \ref{sec:diskstructure} to initialize our simulation.
We used an isothermal equation of state and the disk temperature was assumed to be vertically isothermal. 
We adopted 256 logarithmically-spaced radial grid cells, 288 uniformly-spaced azimuthal grid cells, and 80 uniformly-spaced meridional grid cells.
A constant $\alpha$ viscosity of $3\times10^{-3}$ was applied. 
We used outflow boundary condition at the radial and meridional boundaries. 
We ran hydrodynamic simulations for 50 companion orbits, varying companion mass, semi-major axis, and orbital eccentricity. 

\begin{figure}
\centering
\includegraphics[width=6in]{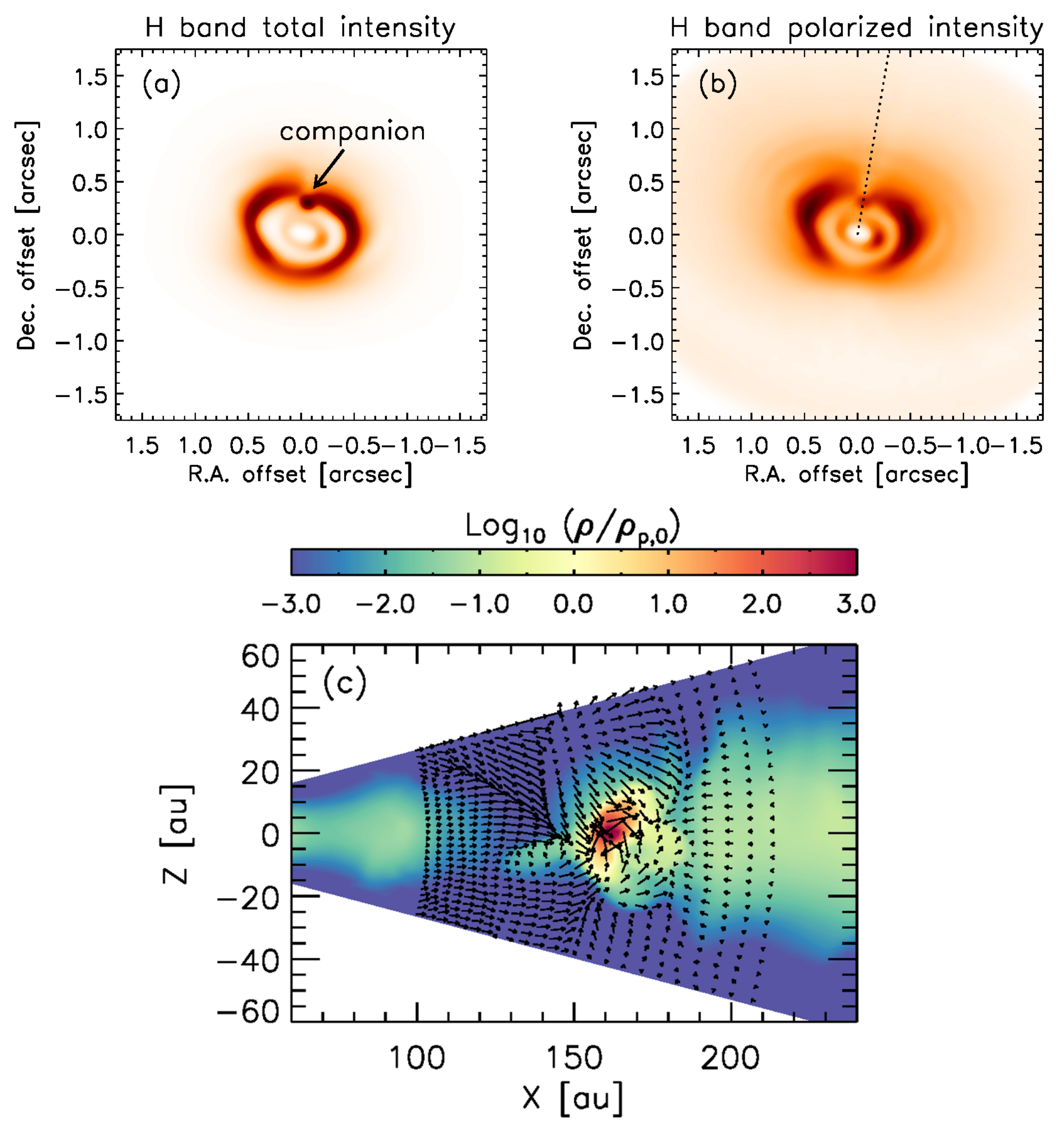}
\caption{Simulated (a) total and (b) polarized intensity maps of HD~34700A in H-band based on the hydrodynamical simulation described in \S\ref{hydro}. The images are smoothed to the same angular resolution as our data. We use a linear color table for the total intensity map and a color table proportional to the square-root of absolute value for the polarized intensity map, consistently with Figures \ref{fig:toti} and \ref{fig:qr}. This model included a planet responsible for carving out the inner cavity and for producing a discontinuity in the ring due to shadowing for the outer disk inner wall by the flows surrounding the planet. Panel (c) presents the disk density distribution along the dotted line crossing the planet's location in panel (b). The density is normalized by its initial value at the location of the planet. The over-plotted arrows present the gas velocity field on the two-dimensional plane, showing complex gas dynamics around the planet. We note that this simulation did not produce the prominent high-pitch-angle spiral arcs seen in the observed images, although some tightly wound outer spiral arms were generated.}
\label{fig:hydro}
\end{figure}

We then generated simulated scattered-light images by post-processing hydrodynamic models using a radiative transfer code, RADMC3D version 0.41\footnote{\url{http://www.ita.uni-heidelberg.de/~dullemond/software/radmc-3d/}}.
We used the same spherical mesh of the hydrodynamic simulations for radiative transfer calculations.
To produce stellar photons we placed two $2~M_\odot$ stars with $3.3~R_\odot$ and $T_{\rm eff} = 5900$K and 5800~K, as constrained above for HD~34700Aa and HD~34700Ab.
For simplicity, we placed both stars at the center of the system. 
We first calculated the dust temperature with a thermal Monte Carlo simulation using $10^8$ photon packages. 
We then computed full polarized scattering off dust particles, using the PA and inclination of the disk obtained in Section \ref{hd34700_spirals}.

Figure~\ref{fig:hydro} shows simulated total and polarized intensity maps in H-band for a 50 $M_J$ planet with semi-major axis of 135 au and orbital eccentricity of 0.2. 
The planet clears the inner disk and generates a ring-like structure beyond its orbit. 
In this model, the violent three-dimensional gas flows around the planet block stellar photons, casting shadows on the ring beyond the planet's location.
This is possible because the gas flows around the planet have a comparable vertical extent to the inner disk and a larger density.
As a result, this shadow produces a feature similar to the observed discontinuity in total and polarized maps.  
We found that the vertical extent and the density of the circumplanetary flows are dependent upon planet mass and orbital eccentricity, and that we need both large planet mass and non-zero eccentricity to reproduce the observed discontinuity (appendix \ref{appendix_hydro}).
We do however caution that our hydrodynamic model may not have a sufficient numerical resolution and proper thermodynamics to accurately simulate the circumplanetary flows.
Future numerical simulations will further test the possibility of circumplanetary flows casting shadows onto the outer disk, and better constrain the mass and orbital eccentricity of the hypothetical planet.

In our hydrodynamic simulations planets with non-zero orbital eccentricity excite multiple spiral arms in the outer disk.
However, these spirals are too tightly wound compared with the observed ones and produce insufficiently strong perturbations, so they are not apparent when the raw images are smoothed to the same angular resolution as the data.
As the pitch angle of and density perturbations driven by spiral arms depend sensitively on both radial and vertical disk thermal structure \citep{zhu15,juhasz18}, future numerical simulations with proper treatments for disk thermodynamics, including heating from spiral shocks, stellar irradiation, and vertical thermal stratification of the disk, will help better understand the origin of the spiral arms in the disk.
From the observational side, better constraints on the disk temperature profile using molecular line emissions will help further examine possible causes of the spirals in the disk.

In general, the inner stellar binary, the planetary orbit, and the disk could all be in somewhat different planes. 
Such inclination angle differences could generate additional dynamics and these interactions should also be explored in future calculations.
Also, given that 50~Jupiter-mass planets/brown dwarfs are rare, we recognize that shadows casted by circumplanetary flows may not be commonly seen.

{Lastly, we comment generally on the how to interpret the shallow, 20--30\arcdeg~ pitch angles observed for the spiral arcs in HD~34700A. Numerical simulations of gravitational instability (GI) in protoplanetary disks show that the pitch angles are typically 10--20\arcdeg \citep{cossins2009,dong2016,forgan2018}. One interesting feature seen in GI simulations is that GI-driven spirals have a constant pitch angle (of course within the same simulation) although why they do so hasn't been understood yet \citep{forgan2018}. 
For companion-driven spirals, pitch angles vary significantly as a function of radius, from $\sim$90 degrees at the vicinity of the companion to almost zero degrees far from the companion  (see Figure 6 of Zhu et al. 2015 and Figure 5 of Bae \& Zhu 2018b). In case of HD 34700B driving the observed spirals, the location where spirals are observed is $\sim$10\% of the distance to HD 34700B (150 vs. 1500 au). For such a situation, the pitch angle is expected to be $<\sim$10 degrees although a warm surface can make spirals more opened in scattered light images \citep[e.g.,][]{benisty2015}.}

\section{Conclusions}

We have presented discovery images of the remarkable disk around HD~34700A, including a large low-density cavity about 1$\arcsec=360$au across. We see signs of possible ongoing planet formation, including a discontinuous ring and a rich series of spiral arcs (possibly up to 8 arcs).  With the new {\em Gaia} distance, we can better identify this system as a young $\sim$5~Myr intermediate-mass binary system (2M$_\odot$+2M$_\odot$) with a very prominent transition disk and not an older debris disk system as previously classified.

Our image analysis and radiative transfer modeling suggest this system is inclined about 42$\arcdeg$ with the North side tilted toward us.  The butterfly pattern in the $U_\phi$ suggests multiple scattering within the disk and this interpretation is supported by our radiative transfer modeling.  This pattern can be a powerful diagnostic of the true geometry of the disk inclination when the disk ring emission is not truly circular.  For inclined optically-thick disks, we caution against using pipeline procedures that aggressively attempt to remove $U_\phi$ signal as a calibration shortcut, without further study as the effect on true $U_\phi$ signals using simulated data.

The brightening of the polarized intensity along the major axis of the ring is reminiscent of Herbig Ae star HD~163296 (MWC 275) disk observed by \citet{garufi2014} and \citet{monnier2017}, but unlike the T Tauri disks of \citet{avenhaus2018} which all show bright $Q_\phi$ on the near-side of the disk, not the edges.  This may mark another difference between Herbigs and T Tauri disks or perhaps it is due to the differences between dust populations that exist at a cavity wall compared to dust existing in the upper layers of a flared disk.  

While the {known} close inner binary can not explain the large transition disk cavity, an inner perturber (i.e., forming exoplanet) can explain the large lower-density hole in the disk as well as some of the inner wall discontinuity.  We found that a sufficiently massive protoplanet could cause local shadowing of the outer disk, reminiscent of the ring ``discontinuity'' we observe for HD~34700A, although shadowing by an inner circumbinary or circumstellar disk could also play a role \citep[as for HD142527;][]{avenhaus2017}. That said, the hydrodynamical simulations predict spiral arms much more tightly wound than observed.  
{Since companion-driven spiral arms are increasingly tightly wound as they propagate \citep{zhu15,baezhu18b} it is difficult to reconcile the observed large pitch angle with the external companion HD 34700B, unless the disk temperature is largely increased (at least at the surface) as suggested for other disks with spirals \citep[e.g.,][]{benisty2015}. } 
Thus, we still lack a definite cause for the multiple spiral arcs for this source in particular and for intermediate-mass stars in general \citep[\`{a} la][]{avenhaus2018}.

Future observations should focus on high angular resolution ALMA gas and dust continuum imaging to continue probing the origin of the spiral structures seen often in intermediate-mass disks.  Also, new visible and near-infrared scattered light imaging with better attention to the absolute photometric calibration will enable next-generation radiative transfer modeling to tightly constrain the dust properties.  Lastly, a search for point sources within the disk might reveal the inner planet responsible for the large cavity seen in the HD~34700A transition disk.

\acknowledgments
J.D.M/A.A. acknowledge support from NSF AST 12100972, 1311698, 1445935, and 1830728. A.~A. also acknowledges support from the CU Boulder Hale Fellowship program.   The RT calculations presented in this paper was performed on the University of Exeter
Supercomputer, a DiRAC Facility jointly funded by STFC, the
Large Facilities Capital Fund of BIS, and the University of Exeter,
and on the DiRAC Complexity system, operated by the University
of Leicester IT Services, which forms part of the STFC DiRAC
HPC Facility (www.dirac.ac.uk). The latter equipment is funded
by BIS National E-Infrastructure capital grant ST/K000373/1 and
STFC DiRAC Operations grant ST/K0003259/1. TJH acknowledges funding from Exeter's STFC Consolidated Grant (ST/J001627/1). SK acknowledges support from an STFC Rutherford Fellowship (ST/J004030/1) and a European Research Council (ERC) Starting Grant (Grant agreement No 639889).
JB acknowledges support from NASA grant NNX17AE31G and computational resources and services provided by the NASA High-End Computing Program through the NASA Advanced Supercomputing Division at Ames Research Center. 

This work has made use of data from the European Space Agency (ESA) mission
{\it Gaia} \citep{gaia2018} (\url{https://www.cosmos.esa.int/gaia}), processed by the {\it Gaia}
Data Processing and Analysis Consortium (DPAC,
\url{https://www.cosmos.esa.int/web/gaia/dpac/consortium}). Funding for the DPAC
has been provided by national institutions, in particular the institutions
participating in the {\it Gaia} Multilateral Agreement.

This research has also made use of the SIMBAD database (\citealt{simbad}), operated at CDS, Strasbourg, France, and
the NASA's Astrophysics Data System Bibliographic Services.

The data presented here were obtained at the Gemini Observatory (programs GS-2017B-LLP-12), which is operated by the Association of Universities for Research in Astronomy, Inc., under a cooperative agreement with the NSF on behalf of the Gemini partnership: the National Science Foundation (United States), National Research Council (Canada), CONICYT (Chile), Ministerio de Ciencia, Tecnolog\'{i}a e Innovaci\'{o}n Productiva (Argentina), Minist\'{e}rio da Ci\^{e}ncia, Tecnologia e Inova\c{c}\~{a}o (Brazil), and Korea Astronomy and Space Science Institute (Republic of Korea).

Lastly, we appreciate feedback from the anonymous referee that led to a more comprehensive and complete article.


%% for the paper.  Each keyword is check against the master list during
%% copy editing.  Individual instruments can be provided in parentheses,
%% after the keyword, but they are not verified.

\vspace{5mm}
\facilities{Gemini:South (GPI)}

\software{IDL, TORUS, FARGO3D, RADMC3D, GPI Data Reduction Pipeline\footnote{http://ascl.net/1411.018}, MPFIT\footnote{http://ascl.net/1208.019}}



\appendix
\section{Definition of Stokes Q,U,Q$_\phi$,U$_\phi$}
\label{appendix:stokes}

\begin{figure}
\centering
\includegraphics[width=6.in]{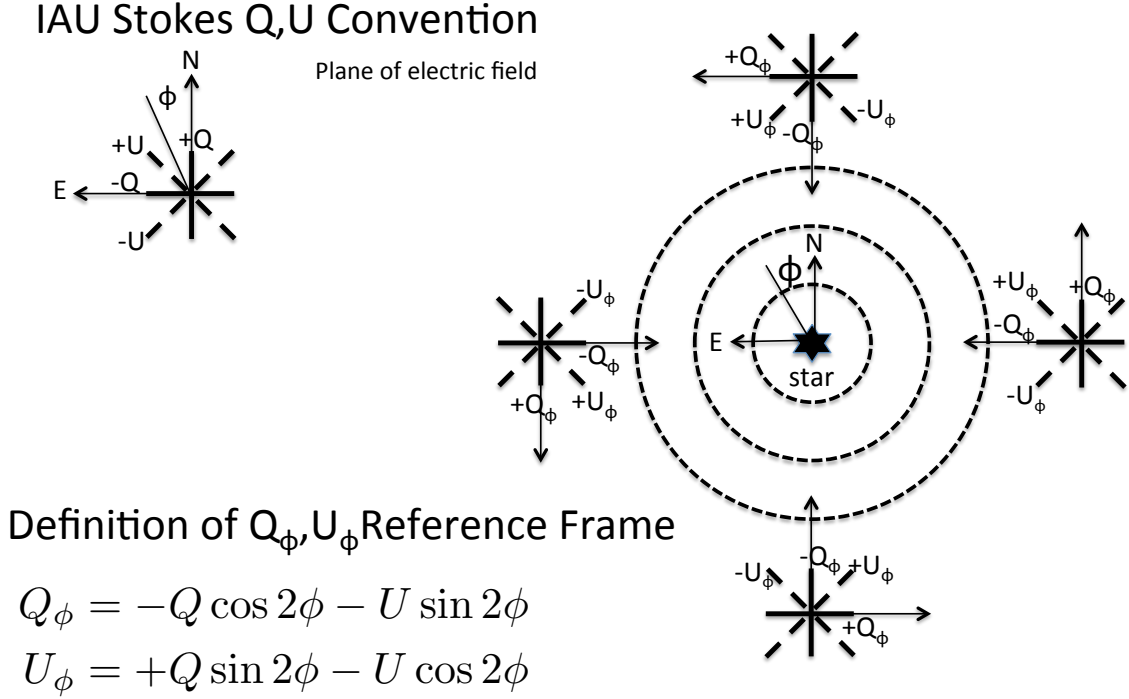}
\caption{Definitions of the Stokes Q,U and $Q_\phi,U_\phi$ used in this paper.} \label{fig:stokes}
\end{figure}

{
We report here for completeness the definition of Stokes $Q,U$ and $Q_\phi,U_\phi$ used in this work.
Figure~\ref{fig:stokes} unambiguously defines our conventions.}

{We have based our convention for $Q,U$ position angles on the IAU standard \citep{iau1974} as summarized by \citet{hamaker1996}.  We have based the definition of $Q_\phi,U_\phi$ on a modified version of the $Q_r,U_r$ system as first proposed by \citet{schmid2006}.  The Schmid definition results in a positive $Q_r$ value for radially-polarized sources, which is not desired for observations of scattered-light disks.  We note that many workers have incorrectly referred to \citet{schmid2006} as the source for their $Q,U,Q_\phi,U_\phi$ convention, but have actually adopted slightly different formulas \citep[e.g.,][]{canovas2015, avenhaus2018}.  
}

{In summary, the Stokes parameters defined in Figure~\ref{fig:stokes} have the following properties which make them attractive:}

\begin{enumerate}
\item {Consistent with IAU recommendations for $Q,U$ position angles}
\item {Yields positive $Q_\phi$ for case when E-field polarization angle at a given pixel is perpendicular to the vector connecting pixel and the star's location}
\item {Results for $Q_\phi,U_\phi$ maps (including the sign $\pm$) agree with previously published polarization imaging by GPI, NACO, and SPHERE groups (despite confusing or incomplete descriptions of conventions contained therein).
}
\end{enumerate}

\section{Hubble Space Telescope Analysis}
\label{appendix_hst}
\begin{figure}
\centering
\includegraphics[width=6.in]{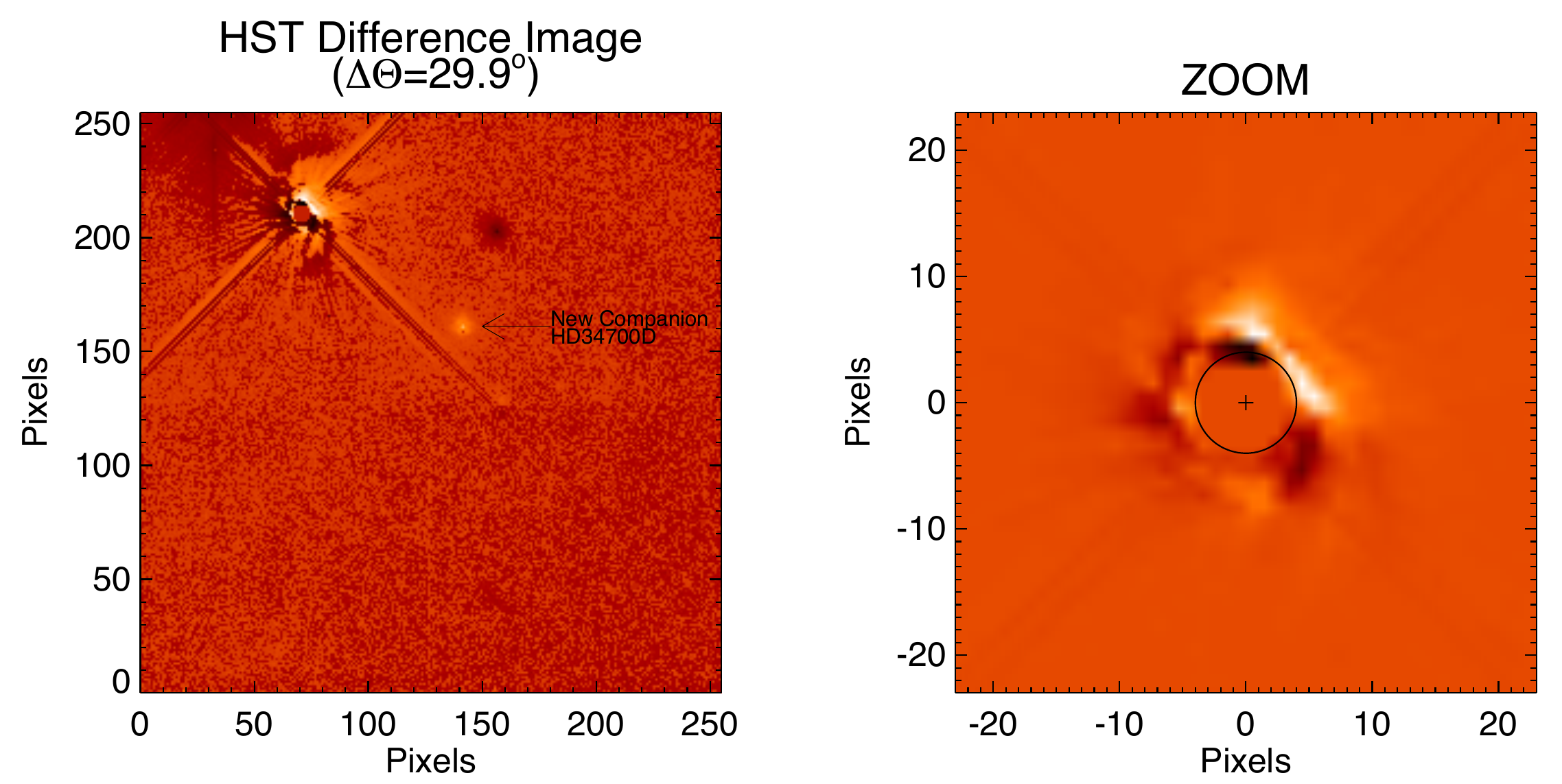}
\caption{Difference of two HST/NICMOS (filter F110W) images with telescope roll angles changed by 29.9$^circ$. (left panel) We see the full image here with the location of the newly discovered HD~34700D marked (separation 6.45'' along PA -60.9\arcdeg). Color table proportional to intensity $|I|^\frac{1}{4}$ while maintaining sign to show low contrast features. (right panel) Here is a zoom-up of the inner 3.5'' around the coronographic spot (marked by black circle) shown with linear intensity color table.   Pixel scale is 75~milliarcseconds/pixel. } \label{fig:hst1}
\end{figure}

The Hubble Space Telescope observed HD~34700A on 1998 Sep 17 using NICMOS/NIC2-CORON (PI: Bradford Smith).  \citet{sterzik2005} already inspected this data to show HD~34700B and HD~34700C were co-moving with HD~34700A.  We were interested in data taken with the wide F110W filter (0.8-1.4$\mu$m) to look for evidence of the scattered light ring we imaged with GPI at a similar wavelength.   Figure\,\ref{fig:hst1} shows the difference between two images taken  at two different telescope roll angles.  In this difference image, the PSF structures should cancel but leave positive/negative imprints of circumstellar structure or companions.  One can see there is complex residual flux around the coronographic spot and what appears to be a newly-discovered companion HD~34700D, a possible fifth  member of the HD~34700 system.  A preliminary analysis indicates that HD~34700D is located at  distance 6.45'' (projected 2300au) along PA -60.9\arcdeg with mag$_{\rm F100W}=$18.6. This flux density corresponds to an absolute magnitude M$_{F110W}$=10.9 leading to a mass estimate of 12$-$15\,M$_J$ \citep[][assuming $t=5$Myr]{chabrier2000}, a borderline planet/brown dwarf object assuming it is physically associated with HD~34700ABC.  New HST data was recently obtained that might allow the proper motion to be determined although these data are still protected at the moment.

\begin{figure}
\centering
\includegraphics[width=6.in]{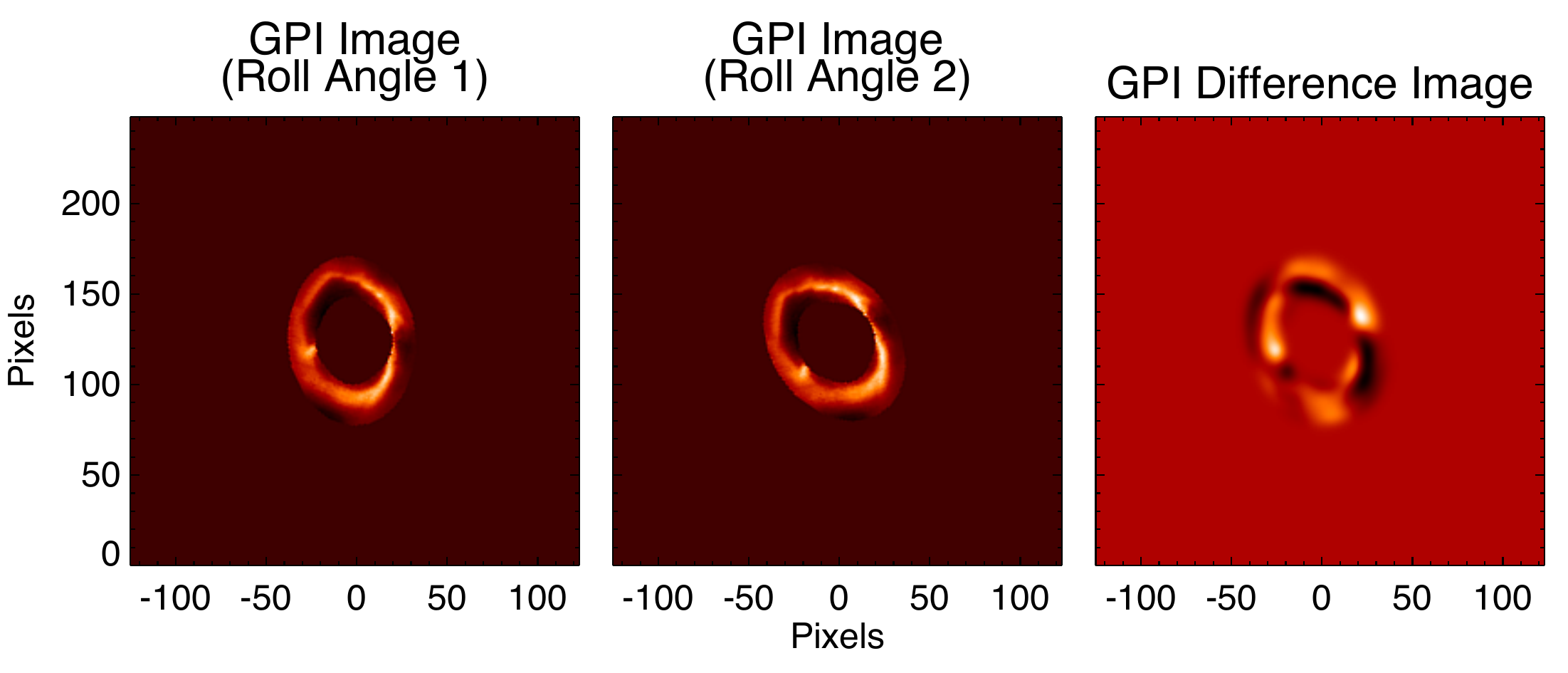}
\caption{This figure shows the GPI total intensity image at the two roll angles observed by HST. The right-most panel shows the difference image and can be directly compared to the right panel in Figure~\ref{fig:hst1}. A mask is applied to only show region where our total intensity image is valid (within 25\% of the elliptical ring structure).  Pixel scale is 14.14~milliarcseconds/pixel and the field-of-view is $\sim$3.5''. } \label{fig:hst2}
\end{figure}

\begin{figure}
\centering
\includegraphics[height=3.in]{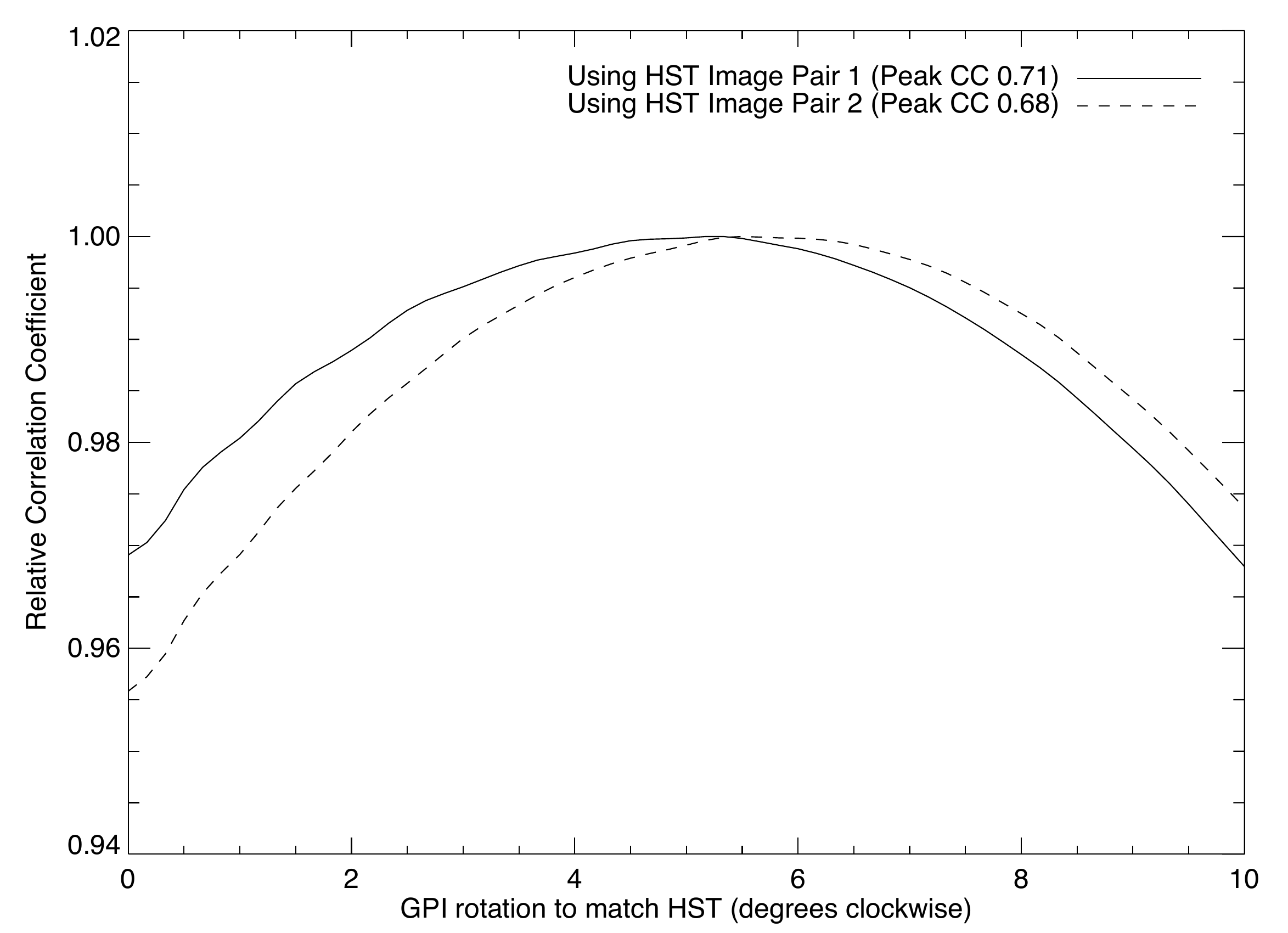}
\caption{{This figure shows the GPI correlation analysis for two independent HST image pairs.  The GPI difference image agrees better with the HST image if we rotate the GPI image by 5.75$\pm$0.25\arcdeg, }suggesting the HD~34700A disk is rotating counter-clockwise with an orbital period of $\sim$1200\,yrs at the radial location of the ring.} \label{fig:hst3}
\end{figure}

In order to understand the residual flux around the coronagraphic spot, we 
simulated the HST roll angle differences using our GPI total intensity image.  Figure~\ref{fig:hst2} shows the GPI image at both roll angles used by HST and the resulting difference.  The correlation between the HST difference image is clear although some differences are apparent.  After a visual inspection, we suspected a very slight rotation between the HST and GPI difference images and we explored this using a correlation analysis over (x,y) shifts and image rotation.  We also varied the size of the occulting mask, smoothing kernal to degrade the GPI resolution to match HST, and also tried using both sets of HST F110W difference images in the archive.  No matter how we changed the details of the correlation analysis, we found the correlation was best ($\sim$71\%) when rotating the GPI image clockwise by $\sim$5.75\arcdeg.  For a 300~mas coronographic spot mask and a GPI smoothing FWHM of 70~mas, Figure~\ref{fig:hst3} shows the correlation coefficient for a range of rotation angles (optimizing the translation match for each candidate rotation angle) with { peak at 5.5\arcdeg for dataset 1 and with a peak at 6.0\arcdeg for dataset 2.  Since uncertainties here are strongly dominated by systematics and not random errors associated with photon noise, we have used these two independent datasets to estimate our optimal rotation angle and associated error:  5.75$\pm$0.25\arcdeg.   Our correlation analysis also allow us to estimate that the GPI flux is $\sim$3.3 higher than the HST data;} while the 19~year time difference and different passbands make a precise cross-calibration difficult, this discrepancy supports our conclusion that the GPI photometry is poorly calibrated in terms of the absolute flux level.

While there is not a large difference in the correlation coefficient between no rotation (68.5\%) and with 5.5\arcdeg rotation (71.2\%), the difference is persistent for multiple datasets and a wide range of methodology details. 
If taken seriously, this would mean the disk is rotating counter-clockwise, consistent with the winding of the spiral arms, with an orbital period of 1200$\pm$50\,yrs. This is in excellent agreement with the expected Keplerian orbital period of 1160\,yr for our disk model (dust ring at 175\,au around a 4\,M$_\odot$ central mass).

As a check, we also performed a more complicated analysis where we first deprojected the ring into a face-on view, applied rotation, then re-projected the result -- this would separate out rotation of the ellipse itself on-sky (not expected) from rotation of structures along the ellipse (expected from orbital motion).  The highest correlation from this analysis was for 8.3$\arcdeg$ rotation, somewhat higher than from our first approach.
Again, we urge caution in interpreting this rotation result since its possible that artifacts from diffraction near the coronographic spot may accidentally mimic the effects of a tiny rotation, but we report the results of our work anyway.

\section{Additional Hydrodynamic Models}
\label{appendix_hydro}
\begin{figure}
\centering
\includegraphics[width=6.in]{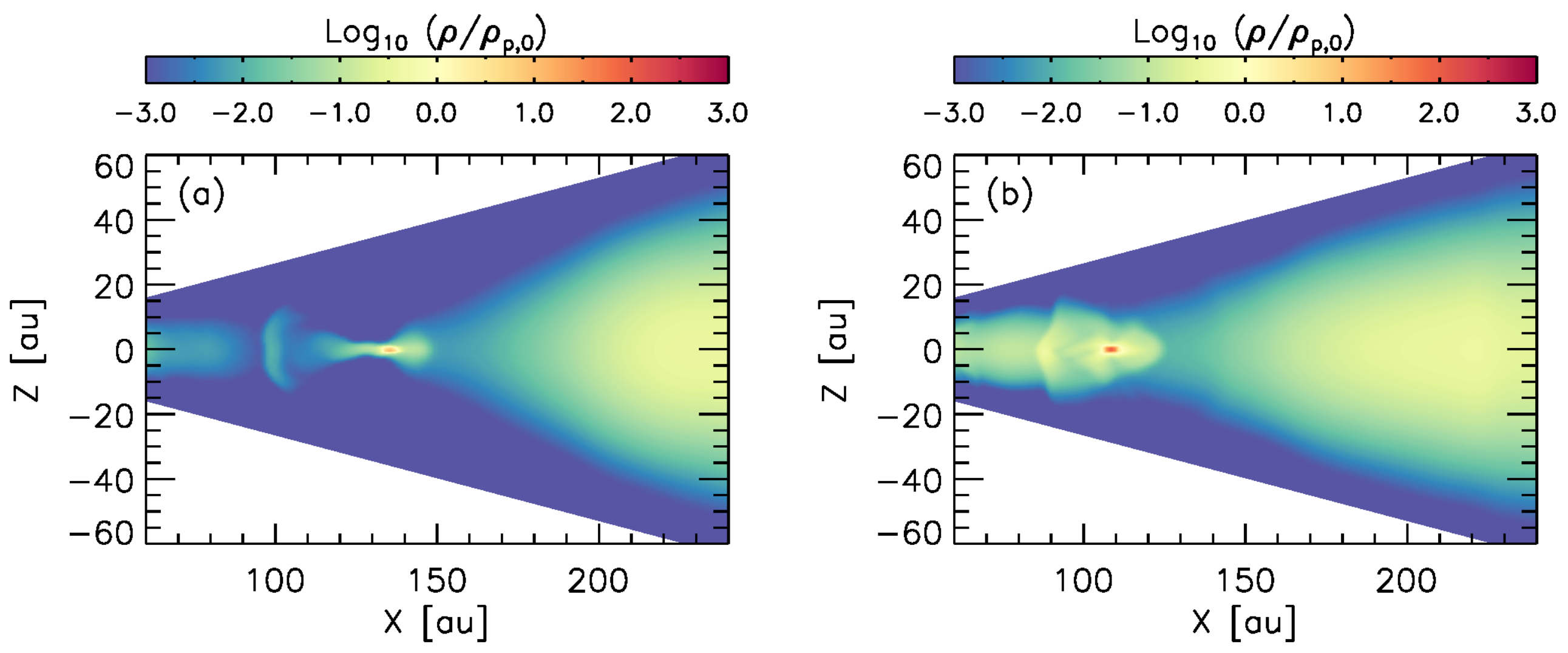}
\caption{Same as Figure \ref{fig:hydro} but with (a) a $50 M_{\rm Jup}$ planet having zero orbital eccentricity, and (b) a $10 M_{\rm Jup}$ planet having 0.2 orbital eccentricity. Compared with the fiducial model presented in Figure \ref{fig:hydro}, these models suggest that a large planetary mass or a non-zero eccentricity alone is not sufficient to create optically-thick three-dimensional flows capable of casting shadow onto the outer disk.} \label{fig:hydro_app}
\end{figure}

The purpose of this appendix section is to make it clear that (1) it is not the circumplanetary disk (CPD) itself creating the shadow shown in Figure~\ref{fig:hydro}, but rather the flow around/onto the CPD; and (2) a hypothetical planet has to have sufficiently large mass and a non-zero eccentricity to produce such flows.
In Figure \ref{fig:hydro_app} we present vertical density distributions from two additional hydrodynamic models. 
Figure \ref{fig:hydro_app}a demonstrates that a 50~M$_J$ planet with a circular orbit has a CPD that remains geometrically-thin.
The planet does not produce vertical flows around it, unlike our fiducial model presented in Section \ref{hydro}.
As the CPD is geometrically-thin, it does not cast a significant shadow onto the outer disk.
Figure \ref{fig:hydro_app}b presents that a 10~M$_J$ planet with 0.2 orbital eccentricity can produce some vertical flows around the CPD.
Comparing with Figure \ref{fig:hydro_app}a this suggests that vertical circumplanetary flows may require a non-zero orbital eccentricity to develop.
In addition, comparing with Figure \ref{fig:hydro}c we find that the strength and vertical extent of circumplanetary flows is dependent on planet mass: a larger planet mass results in stronger and more vertically extended flows.
When post-processed with radiative transfer calculations, however, this model fails to reproduce the observed discontinuity in the outer ring, presumably because the density and/or vertical extent of the circumplanetary flows is not sufficient. 

\section{Additional Figures}
\begin{figure}
\centering
\includegraphics[height=3.in]{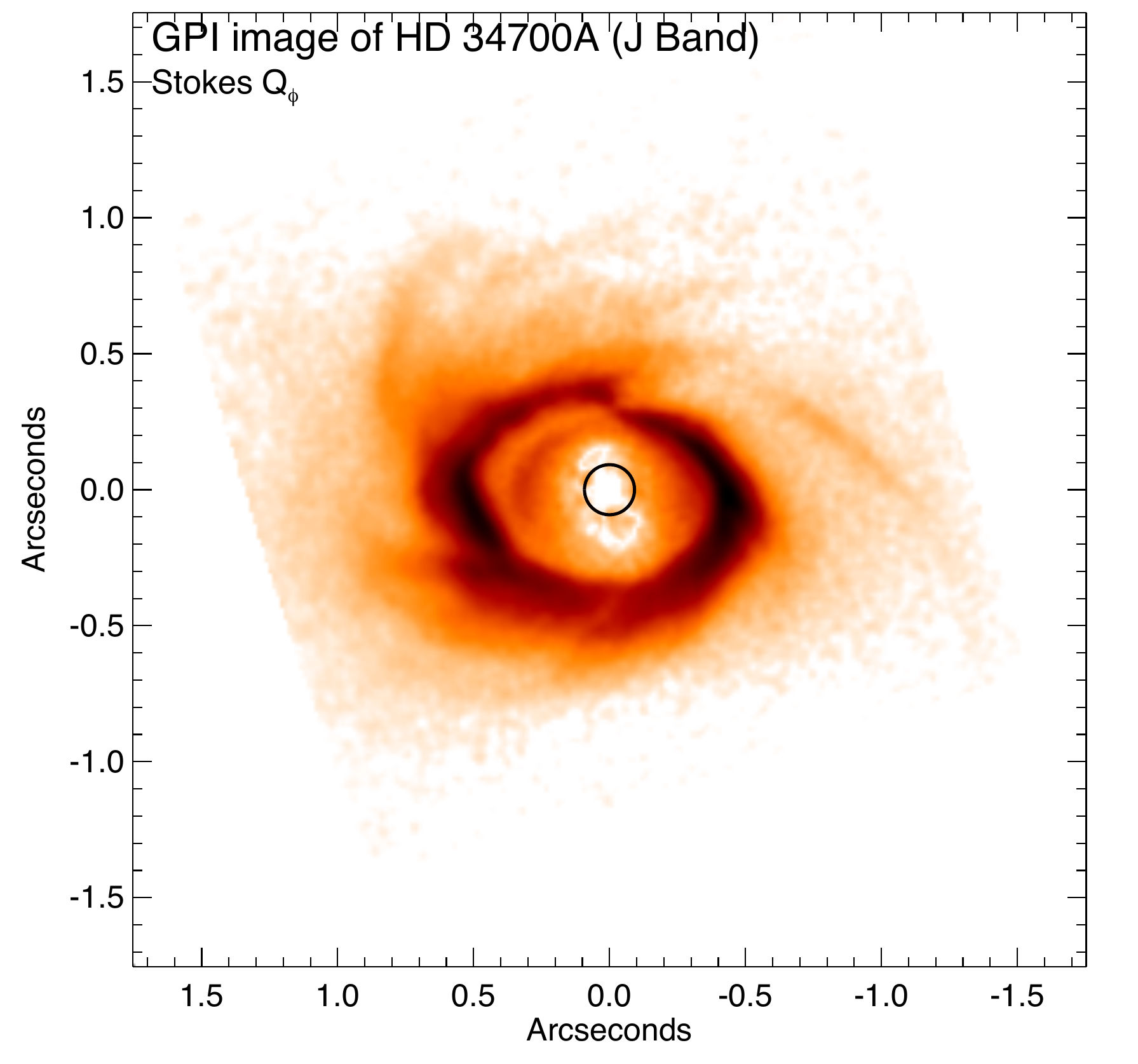}
\hphantom{........}
\includegraphics[height=3.in]{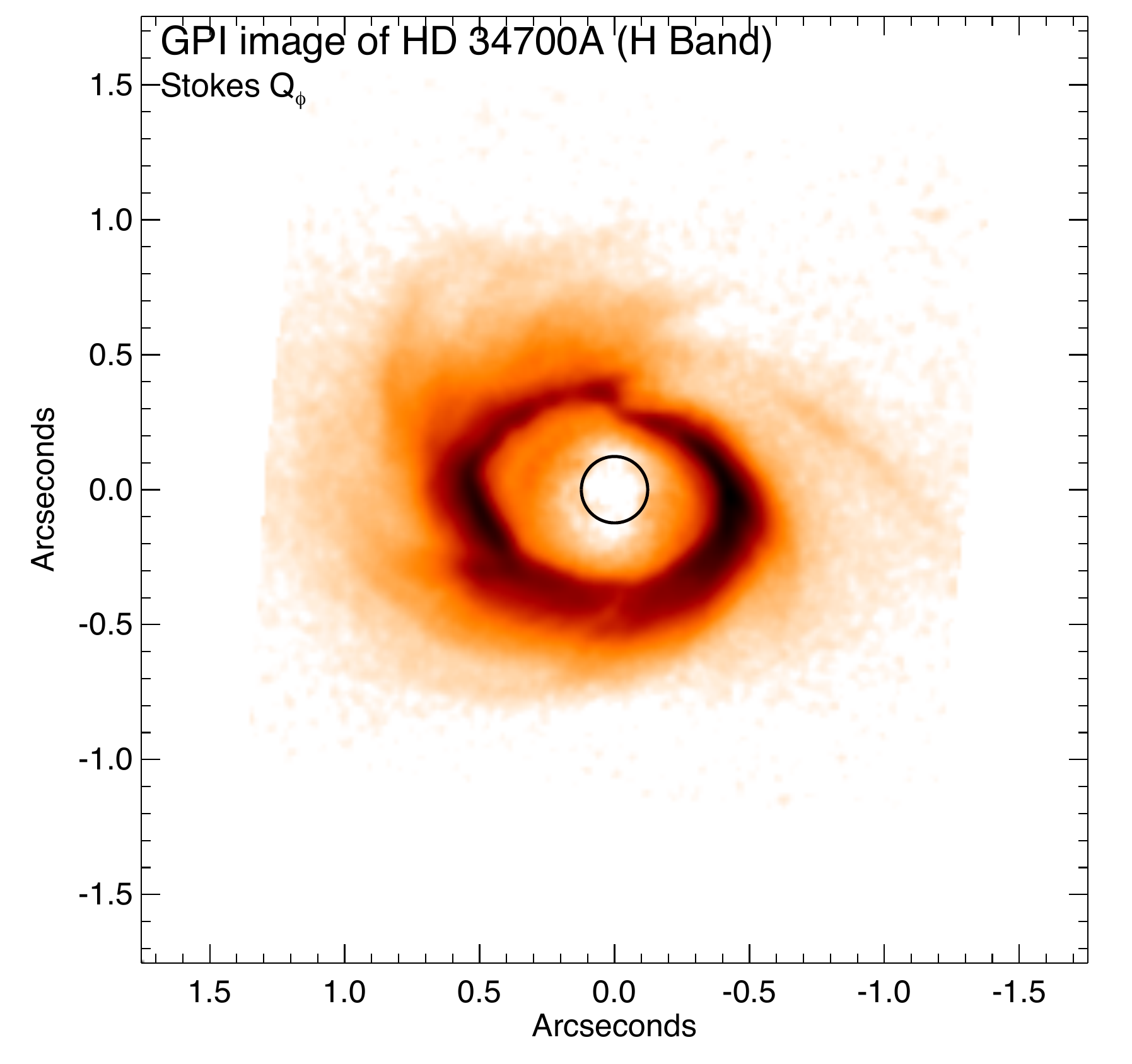}
\caption{The polarized intensity $Q_\phi$ images of HD~34700A in J band (left panel) and in H band (right panel). We present these images without contours  to aid researchers in comparing our results with multi-wavelength imaging data from other facilities -- see Figure~\ref{fig:qr} for full details on the color table. The circle marks the location and size of the coronagraphic spot used for the H-band  observations.}
\label{niceimages1}
\end{figure}

\begin{figure}
\centering
\includegraphics[height=6.in]{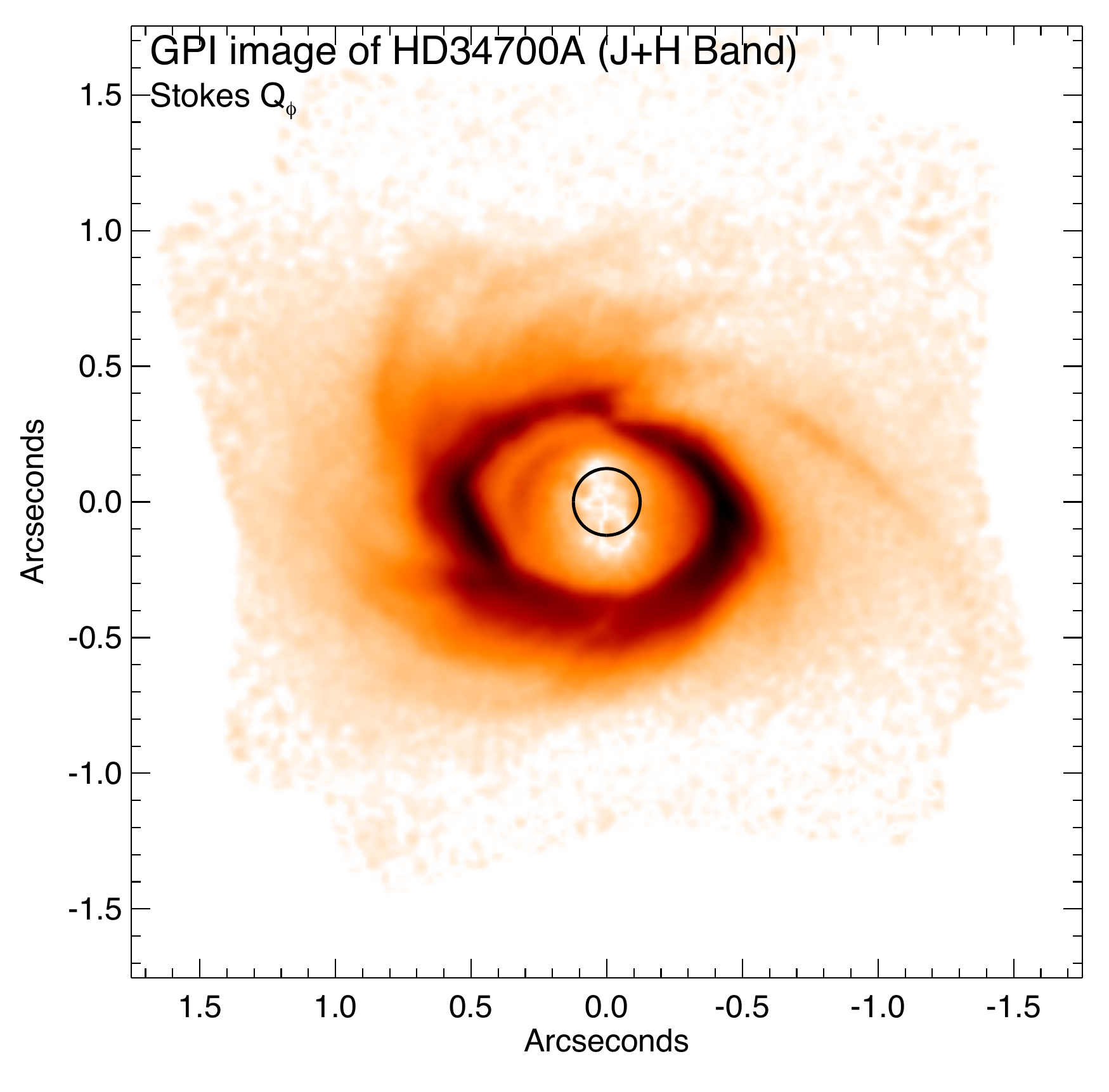}

\caption{We present a combined J-band and H-band polarized intensity image to achieve best signal-to-noise ratio for identifying features in the disk.   These images are shown without contours  to aid researchers in comparing our results with multi-wavelength imaging data from other facilities. The circle marks the location and size of the coronagraphic spot used for the H-band observations. } \label{niceimages2}
\end{figure}

In order to aid other researchers in comparing our results to images taken at other wavelengths, we provide  reference figures (see Figures~\ref{niceimages1} \& \ref{niceimages2}) here of our polarized intensity $Q_\phi$ surface brightness maps without contours or distracting labels.

\bibliographystyle{aasjournal}
\bibliography{monnier_gpi}

\listofchanges

\end{document}